\documentclass[iop]{emulateapj}
\usepackage{amsmath, amsthm}
\usepackage{booktabs}

\usepackage{comment}

\slugcomment{Draft, \today}

\newcommand*{\Ledd}{L_{\rm Edd}}
\newcommand*{\calM}{{\cal M}}
\newcommand*{\calFg}{{\cal F}_{\rm g}}

\newcommand*{\Mdotb}{\dot M_{\rm B}}
\newcommand*{\Mdotbe}{\dot M_{\rm e}}

\newcommand*{\Mdott}{\dot M_{\rm t}}

\newcommand*{\lambdat}{\lambda_{\rm t}}

\newcommand*{\Lambdacr}{\Lambda_{\rm cr}}
\newcommand*{\lambdacr}{\lambda_{\rm cr}}

\newcommand*{\gmin}{g_{\rm min}}
\newcommand*{\fmin}{f_{\rm min}}
\newcommand*{\xmin}{x_{\rm min}}

\newcommand*{\Mbh}{M_{\rm BH}}
\newcommand*{\Mg}{M_{\rm g}}
\newcommand*{\rhog}{\rho_{\rm g}}

\newcommand*{\cs}{c_{\rm s}}
\newcommand*{\rb}{r_{\rm B}}
\newcommand*{\rbe}{r_{\rm e}}
\newcommand*{\rg}{r_{\rm g}}
\newcommand*{\rmin}{r_{\rm min}}

\newcommand*{\rhoinf}{\rho_{\infty}}
\newcommand*{\pinf}{p_{\infty}}
\newcommand*{\csinf}{c_{\infty}}
\newcommand*{\Tinf}{T_{\infty}}
\newcommand*{\rhotil}{\tilde\rho}
\newcommand*{\mpr}{m_{\rm p}}

\newcommand\rhos{\rho_*}

\newcommand\rhoDM{\rho_{\rm DM}}
\newcommand\rhon{\rho_{\rm n}}

\newcommand\rhoNFW{\rho_{\rm NFW}}

\newcommand\rs{r_*}

\newcommand\reff{R_{\rm e}}
\newcommand\ra{r_{\rm a}}
\newcommand\rt{r_{\rm t}}

\newcommand\Phig{\Phi_{\rm g}}

\newcommand\Psin{\Psi_{\rm n}}

\newcommand\Ms{M_*}

\newcommand\MR{{\cal R}}
\newcommand\MRg{{\cal R}_{\rm g}}
\newcommand\Rm{{\mathcal{R}}_{\rm m}}

\newcommand\Ks{K_*}

\newcommand\Wg{W_{\rm *g}}
\newcommand\Wgt{{\widetilde W}_{\rm g}}
\newcommand\Wbh{W_{\rm *BH}}

\newcommand\sigv{\sigma_{\rm V}}
\newcommand\sigp{\sigma_{\rm p}}
\newcommand\srad{\sigma_{\rm r}}
\newcommand\sigBH{\sigma_{\rm BH}}
\newcommand\sigg{\sigma_{\rm g}}
\newcommand\sigBHp{\sigma_{\rm BHp}}
\newcommand\siggp{\sigma_{\rm gp}}

\newcommand\Tv{T_{\rm V}}

\newcommand\MD{M_{\rm {DM}}}

\newcommand\csig{\xi_{\rm g}}
\newcommand\csih{\xi_{\rm NFW}}
\newcommand\Rinf{R_{\rm inf}}
\newcommand\betac{\beta_{\rm c}}

\shortauthors{L. Ciotti and S. Pellegrini}
\shorttitle{Isothermal Bondi accretion in two-component Jaffe galaxies}

\begin{document}

\title{Isothermal Bondi accretion in two-component Jaffe galaxies with a central black hole}
\author{Luca Ciotti$^{\star}$ and Silvia Pellegrini}
\affil{Department of Physics and Astronomy, University of Bologna,
  via Piero Gobetti 93/2, 40129 Bologna, Italy\\
$^{\star}$E-mail: luca.ciotti@unibo.it }

%\date{\today}

\begin{abstract}

  The fully analytical solution for isothermal Bondi accretion on a
  black hole (MBH) at the center of two-component Jaffe (1983) galaxy
  models is presented. In a previous work we provided the analytical
  expressions for the critical accretion parameter and the radial
  profile of the Mach number in the case of accretion on a MBH at the
  center of a spherically symmetric one-component Jaffe galaxy model.
  Here we apply this solution to galaxy models where both the stellar
  and total mass density distributions are described by the Jaffe
  profile, with different scale-lengths and masses, and to which a
  central MBH is added. For such galaxy models all the relevant
  stellar dynamical properties can also be derived analytically
  (Ciotti \& Ziaee Lorzad 2018). In these new models the
  hydrodynamical and stellar dynamical properties are linked by
  imposing that the gas temperature is proportional to the virial
  temperature of the galaxy stellar component.  The formulae that are
  provided allow to evaluate all flow properties, and are then useful
  for estimates of the scale-radius and the mass flow rate when
  modeling accretion on massive black holes at the center of galaxies.
  As an application, we quantify the departure from the true mass
  accretion rate of estimates obtained using the gas properties at
  various distances from the MBH, under the hypothesis of classical
  Bondi accretion.

\end{abstract}


\section{Introduction} 

Observational and numerical investigations of accretion on massive
black holes (hereafter MBH) at the center of galaxies often lack the
resolution to follow gas transport down to the parsec scale. In these
cases, the {\it classical} Bondi (1952) solution for
spherically-symmetric, steady accretion of a spatially infinite gas
distribution onto a central point mass is then commonly
adopted; this is the standard reference for estimates of the accretion radius
(i.e., the sonic radius), and the mass accretion rate (see, e.g., Rafferty et al. 2006,
Sijacki et al. 2007; Di Matteo et al. 2008; Gallo et al. 2010; Pellegrini 2010; 
Barai et al. 2011; Bu et al. 2013; Cao 2016; Volonteri et al. 2015; Choi et al. 2017; Park et
al. 2017; Beckmann et al. 2018; Ram\'irez-Velasquez et al. 2018; Barai et al. 2018).
Even though highly idealized, during phases of moderate accretion (in
the ``maintainance'' mode), indeed, the problem can be considered
almost steady, and Bondi accretion could provide a reliable
approximation of the real situation (e.g., Barai et al. 2012, Ciotti \& Ostriker 2012).

However, leaving aside the validity of the fundamental assumptions of
spherical symmetry, stationarity, and optical thinness, two major
problems affect the direct application of the classical Bondi solution, namely the facts that 1)
the boundary values of density and temperature of the accreting gas
should be evaluated at infinity, and 2) in a galaxy, the gas experiences the
gravitational effects of the galaxy itself (stars plus dark matter)
all the way down to the central MBH, and the MBH gravity becomes
dominant only in the very central regions, inside the so-called
``sphere-of-influence''.  The solution commonly adopted in numerical
and observational applications to alleviate these problems is to use
values of the gas density and temperature ``sufficiently near'' the
MBH, thus assuming that the galaxy effects are negligible. Of
course, as the density and temperature of the accreting gas change along
the pathlines, also the predictions of classical Bondi accretion will
change, when based on the density and temperature measured at a finite
distance from the MBH. It is therefore of great interest to be
able to quantify the systematic effects on the accretion radius and the
mass accretion rate obtained
from the classical Bondi solution, due to measurements taken at finite distance from
the MBH, and under the effects of the galaxy potential well.

A first step towards a quantititative analysis of this problem was
carried out in Korol et al. (2016, hereafter KCP16) where 
the Bondi problem was generalized to the case of mass accretion at the
center of galaxies, including also the effect of electron scattering
on the accreting gas. KCP16 then calculated the deviation from the true
values of the estimates of the Bondi radius and of the mass accretion
rate, due to adopting as boundary values for the density and
temperature those at a finite distance from the MBH, and assuming the
validity of the classical Bondi accretion solution. In the specific
case of Hernquist (1990) galaxies, KCP16 obtained the analytical
expression of the critical accretion parameter, as a function of the
galaxy properties and of the gas polytropic index $\gamma$. However,
even for this quite exceptional case, the radial profiles of the
hydrodynamical variables remained to be determined
numerically. Following KCP16,  Ciotti \& Pellegrini (2017, hereafter
CP17) showed that the whole accretion solution can be given in an
analytical way (in terms of the Lambert-Euler $W$-function) for the
{\it isothermal} accretion in Jaffe (1983) and Hernquist galaxy models
with central MBHs.  This meant that for these models not only it is
possible to express analytically the critical accretion parameter, but
also that the whole radial profile of the Mach number (and then of all
the hydrodynamical functions) can be explicitely written.  At the best
of our knowledge, CP17 provided the first fully analytical solution of
the accretion problem on a MBH at the center of a galaxy.

The galaxy models used in KCP16 and CP17, i.e., the Hernquist and
Jaffe models, are not only relevant because for them it is possible to
solve the accretion problem, but also because of their numerous
applications in Stellar Dynamics. In fact, these models belong to the
family of the so-called $\gamma$-models (Dehnen 1993, Tremaine et
al. 1994) and are known to reproduce very well the radial trend of the
stellar density distribution of real elliptical galaxies; at the same
time, their simplicity allows for analytical studies of one and
two-component galaxy models (e.g., Carollo et al. 1995, Ciotti et
al. 1996, Ciotti 1999). In particular, Ciotti \& Ziaee Lorzad (2018,
hereafter CZ18), expanding a previous study by Ciotti et al.  (2009),
presented spherically symmetric two-component galaxy models (hereafter
JJ models), where the {\it stellar} and {\it total} mass density
distributions are both described by the Jaffe profile, with different
scale-lengths and masses, and a MBH is added at the center. The
orbital structure of the stellar component is described by the
Osipkov-Merritt anisotropy (Merritt 1985). Moreover, the dark matter
halo (resulting from the difference between the total and the stellar
distributions) can reproduce remarkably well the Navarro et al. (1997;
hereafter NFW) profile, over a very large radial range, and down to
the center. Among other properties, for the JJ models the solution of
the Jeans equations and the relevant global quantities entering the
Virial Theorem can be expressed analytically. Therefore, the JJ models
offer the {\it unique} opportunity to have a simple yet realistic
family of galaxy models with a central MBH, allowing both for the
fully analytical solution of the Bondi (isothermal) accretion problem
{\it and} for the fully analytical solution of the Jeans equations;
all this permits then a simple joint study of stellar dynamics and
fluidodynamics without resorting to ad-hoc numerical codes.

This paper is organized as follows.  In Section 2 we recall the main
properties of the Jaffe isothermal accretion solution, and in Section
3 we list the main properties of the JJ models. In Section 4 we show
how the structural and dynamical properties of the stellar and dark
matter components can be linked to the parameters appearing in the
accretion solution. In Section 5 we examine the departure of the
estimate of the mass accretion rate from the true value, when the
estimate is obtained using as boundary values for the density and
temperature those at points along the solution, at finite distance
from the MBH. The main conclusions are summarized in Section 6.

%%%%%%%%%%%%%%%%%%%%%%%%%%%%%%%%%%%%%%%%%%%%%%%%%%%%%%%%%%%%%%%%%
\section{Isothermal Bondi Accretion in Jaffe Galaxies with a Central
  MBH, and with electron scattering}
\label{sec:class}

Following KCP16, and in particular the full treatment of the
isothermal case in CP17, we shortly recall here the main properties of 
isothermal Bondi accretion, in the potential of a Jaffe galaxy
hosting a MBH at its center. We begin with the classical Bondi case.

\subsection{The classical Bondi solution}

In the classical Bondi problem, the gas is perfect, has a spatially
infinite distribution, and is accreting on to a MBH, of mass
$\Mbh$. The gas density and pressure are linked by the polytropic
relation
\begin{equation} 
p = {k_{\rm B} \rho T\over <\mu>\mpr} = \pinf\rhotil^{\gamma},\quad
\rhotil\equiv{\rho\over\rhoinf},
\end{equation}
where $\gamma$ is the polytropic index ($\gamma =1$ in the isothermal
case), $\mpr$ is the proton mass, $<\mu>$ is the mean molecular
weight, $k_{\rm B}$ is the Boltzmann constant, and $\pinf$ and
$\rhoinf$ are respectively the gas pressure and density at infinity.
The sound speed is $\cs=\sqrt{\gamma p/\rho}$, and of course in the
isothermal case it is constant, $\cs=\csinf$, its value at infinity.

The time-independent continuity equation is:
\begin{equation} 
4 \pi r^2 \rho(r) v(r)= \Mdotb,
\end{equation}
where $v(r)$ is the modulus of the gas radial velocity, and $\Mdotb$ is the
time-independent accretion rate on the MBH.  An important scalelength
of the problem, the so-called Bondi radius, is naturally defined as
\begin{equation} 
\rb \equiv {G\Mbh\over\csinf^2}:
\end{equation}
we stress that the Bondi radius remains defined by
the equation above {\it independently} of the presence of the galactic
potential. After introducing the normalized quantities
\begin{equation}
x\equiv {r\over\rb},\quad
\calM(r)={v(r)\over\cs(r)},
\end{equation}
where $\calM$ is the Mach number, eq. (2) determines the accretion
rate for assigned $\Mbh$ and boundary conditions:
\begin{equation}
x^2 \rhotil (x)\calM (x) ={\Mdotb\over 4\pi\rb^2\rhoinf\csinf}\equiv\lambda,
\end{equation}
where $\lambda$ is the dimensionless accretion parameter.
In the isothermal case, the classical
Bondi problem (e.g., KCP16) reduces to the solution of the following system:
\begin{equation}
\begin{cases} 
g(\calM)= f(x) -\Lambda,\qquad\Lambda \equiv \ln\lambda,
\\ \\
\displaystyle{g(\calM)\equiv {\calM^2\over 2}-\ln\calM ,}
\\ \\
\displaystyle{f(x)\equiv {1\over x}+2\ln x}.
\end{cases}
\end{equation}
As well known, $\Lambda$ cannot be chosen arbitrarily; in fact, both
$g(\calM)$ and $f(x)$ have a minimum, and 
\begin{equation}
\begin{cases}
\displaystyle{\gmin={1\over 2}, \qquad\qquad \quad {\rm for}\quad\quad \calM_{\rm min}=1,}
\\ \\
\displaystyle{\fmin= 2-2\ln 2, \qquad {\rm for} \quad\quad \xmin={1\over 2}.}
\end{cases}
\end{equation}
Solutions of eq. (6) exist only for $\gmin \le \fmin -\Lambda$,
i.e., for $\Lambda\leq\Lambdacr\equiv\fmin-\gmin$, which in turn 
is equivalent to the condition
\begin{equation}
\lambda \leq \lambdacr = {{\rm e}^{3/2}\over 4}.
\end{equation}
Along the {\it critical solutions}, i.e., the solutions of eq. (6) for
$\lambda =\lambdacr$, $\xmin$ marks the position of the {\it sonic
  point}, i.e., $\calM (\xmin) =1$. For $\lambda <\lambdacr$ instead
two regular subcritical solutions exist, one everywhere subsonic and
one everywhere supersonic, with the respective maximum and the minimum
value of $\calM (x)$ reached at $\xmin$.

Summarizing, the solution of the classical Bondi problem requires to
determine $\xmin$ and so $\lambdacr$, and possibly to obtain the
radial profile $\calM(x)$ for given $\lambda \le \lambdacr$ (see,
e.g., Bondi 1952; Frank, King \& Raine 1992, KCP16, CP17). Once
$\lambda$ is assigned and $\calM (x)$ is known, all functions involved
in the accretion problem are known from eqs. (5) and (1): for example,
along the critical solution,
\begin{equation}
\rhotil(x)={\lambdacr\over x^2\calM (x)},
\end{equation}
while the modulus of the accretion velocity in the isothermal case 
is $v(r)=\csinf\calM(x)$. 

\subsection{Isothermal Bondi accretion (with electron scattering) in Jaffe galaxies}

The classical Bondi problem can be generalized by including the
effects of radiation pressure due to electron scattering, and the
additional gravitational field of the host galaxy.  In fact, the
accretion flow can be affected by the emission of radiation near the
MBH, that exerts an outward pressure (see, e.g., Mo\'scibrodzka \& Proga 2013
for a study of the irradiation effects on the flow). In the optically thin case the
radiation feedback is implemented as a reduction of the gravitational
force of the MBH, by the factor
\begin{equation}
       \chi\equiv 1- {L\over\Ledd},
\end{equation}
where $L$ is the accretion luminosity, $\Ledd=4\pi c G \Mbh\mpr
/\sigma_{\rm T}$ is the Eddington luminosity, and $\sigma_{\rm T}=6.65
\times 10^{-25}$cm$^2$ is the Thomson cross section. Note that the
maximum luminosity remains equal to $\Ledd$ as defined above even in
presence of the potential of the host galaxy.  As shown in KCP16 and
CP17 (see also Lusso \& Ciotti 2011), for the Bondi problem on an
isolated MBH the critical value $\lambdacr$ and the mass accretion
rate modified by electron scattering can be calculated analytically,
with the new critical parameter given by $\chi^2\lambdacr$.

The more general problem of
Bondi accretion with electron scattering in the potential of a galaxy
hosting a central MBH was addressed in KCP16 and CP17. In particular,
it was shown that there is an analytical expression for $\xmin$ in the
isothermal case in Jaffe galaxies with a central MBH, and in the
generic polytropic case for Hernquist galaxies with a central MBH; thus,
in both cases it is possible to determine, also in presence of
radiation pressure,  the value of the critical accretion parameter
(that now we call $\lambdat$). For Hernquist galaxies, the polytropic
problem leads to the solution of a cubic equation, producing one or
two sonic points (depending on the specific choice of the galactic
parameters), while in the isothermal Jaffe case the relevant equation
is quadratic, and only one sonic point exists, independently of the
galaxy parameters. In addition, CP17 showed that isothermal Bondi
accretion cannot be realized in Hernquist galaxies with $\Mbh =0$ (or
$\chi =0$) while it is possible in a subset of Jaffe galaxies
(provided a simple inequality is satisfied among the galaxy
parameters). Summarizing, since also $\calM$ is given analytically in
the isothermal case, a fully analytical solution exists for isothermal
accretion on MBHs at the center of Hernquist and Jaffe galaxies.
However, due to the complicacies of Bondi accretion in Hernquist
galaxies, and given that two-component JJ models with a total Jaffe
potential and a central MBH have been recently presented (CZ18), in
the rest of the paper we restrict to the case of two-component Jaffe
galaxies. Of course, the existence of the analytical accretion
solution for the one-component Hernquist model with central MBH
guarantees that a similar analysis could be done for the two-component
Hernquist analogues of JJ models.

In the remainder of the Section we recall the main properties of isothermal Bondi accretion in a
Jaffe total potential with a central MBH (CP17); in Section 4, these will be
used to address the problem of accretion in JJ models.  The
gravitational potential of a Jaffe density distribution of
total mass $\Mg$ and scale-length $\rg$ is given by:
\begin{equation} 
	\Phig={G\Mg\over\rg}\ln {r\over r+\rg},
\end{equation}
and, with the introduction of the two parameters:
\begin{equation} 
\MR \equiv {\Mg\over\Mbh},\quad 
\xi \equiv {\rg\over\rb},
\end{equation}
the function $f$ in eq. (6) becomes: 
\begin{equation} 
	f = {\chi\over x}  - {\MR\over\xi}\ln {x\over x+\xi} +  2\ln
        x,\quad
x\equiv{r\over\rb}.
\end{equation}
Note how, for $\MR\to 0$ (or $\xi\to\infty$), the galaxy
contribution vanishes, and the problem reduces to the classical
case. In the limit of $\chi=0$ ($L=\Ledd$)\footnote{Due to a typo, 
before their Sect. 4.1, KCP16 wrote that this case corresponds to $\chi=1$.}, radiation pressure cancels exactly the
MBH gravitational field, and the problem describes
accretion in the potential of the galaxy only, without electron
scattering and MBH.  When $\chi=1$ ($L=0$), the radiation pressure has
no effect on the accretion flow.

The presence of the galaxy potential and electron scattering changes
the accretion rate, that (in the critical case) we now indicate as
\begin{equation} 
	\Mdott = 4 \pi \rb^2 \lambdat \rhoinf 
      \csinf={\lambdat\over\lambdacr}\Mdotb,
\end{equation}
where again $\Mdotb$ is the classical critical Bondi accretion
rate for the same chosen boundary conditions $\rhoinf$ and $\csinf$ in
eq. (5), and $\lambdat$ is the critical accretion parameter of the new
problem. From the same arguments presented in Sect. 2.1, $\lambdat$ is
known once the absolute minimum $\fmin(\chi,\MR, \xi)$ is known;
this, in turn, requires the determination of $\xmin (\chi,\MR,\xi)$,
while the function $g(\calM)$ is unaffected by the addition of the
galaxy potential.

As shown in CP17 for the Jaffe galaxy, the position of the only minimum
of $f$ (corresponding to the sonic radius of the critical
solution) in eq. (13) is given by:
\begin{equation}
\xmin\equiv {\rmin\over\rb}= {\MR + \chi -2\xi + \sqrt{(\MR +\chi -2\xi)^2 + 8\chi\xi}\over 4},
\end{equation}
and then one can evaluate $ \fmin =f(\xmin)$ and $\ln\lambdat = \fmin -
\gmin$. In the peculiar case of $\chi =0$, a solution of the accretion
problem is possible only for $\MR\geq 2\xi$, with:
\begin{equation}
\xmin ={\MR -2\xi\over 2},\quad 
\lambdat ={\MR^2\over 4\sqrt{e}}\left(1-{2\xi\over\MR}\right)^{2-\MR/\xi}. 
\end{equation}
When $\MR\to2\xi$, then $\xmin \to 0$ (the sonic point is at the
origin), $\fmin\to 2\ln\xi$, and finally $\lambdat \to
\xi^2/\sqrt{e}$. Note that the $\chi=0$ case can be {\it also}
interpreted as the case of a null MBH mass, and the formulae in
eq. (16) can be used provided the dependence of $\MR$, $\xi$ and
$\rb$ on $\Mbh$ is factored out, and simplified before considering the
limit for $\Mbh\to 0$. Thus, when $\Mbh =0$ the condition for the
existence of the solution, and the position of the sonic radius, are:
\begin{equation}
{\MR\over 2\xi}={G\Mg\over 2\rg\csinf^2}\geq 1,\quad
{\rmin\over\rg}={G\Mg\over 2\csinf^2\rg}-1.
\end{equation}
Moreover, even though $\lambdat$ in eq. (16) diverges for $\Mbh\to 0$,
the accretion rate given in eq. (14) remains finite also in this case,
with a value that can be easily calculated in closed form in terms of
the galaxy parameters.

The radial trend of the Mach number for the critical accretion
solution of eq. (6) with $f$ in eq. (13), and $\lambda =\lambdat$,
is given by eq. (35) in CP17, that is:
\begin{equation}
\calM^2= -
\begin{cases}
\displaystyle{
W\left(0, - {\rm e}^{-2f }\lambdat^2\right), \quad x\geq \xmin ,
}
\\ \\
\displaystyle{
W\left(-1, - {\rm e}^{-2f}\lambdat^2\right), 
\quad\quad 0< x\leq \xmin,
}
\end{cases}
\end{equation}
where $W$ is the Lambert-Euler function, and its relevant properties
are given in CP17 (Appendix A, see also Appendix B in CZ18).  Note
that the subcritical ($\lambda<\lambdat$) solutions are obtained by
using $W(0,z)$ for the subsonic branch, and $W(-1,z)$ for the
supersonic branch. It is useful to recall that from the expansion for
$x\to 0^+$ of the supersonic branch of $W(-1,z)$ in eq. (18), one has
that for $\chi >0$, $\calM (x)\sim {\sqrt{2\chi}}x^{-1/2}$, while for
$\chi =0$, $\calM (x)\sim 2{\sqrt{(1-\MR/2\xi)\ln(x/\xmin)}}$
(provided that ${\cal R} \geq 2\xi$); moreover, the expansion of
$\calM(x)$ for $x \to \infty$ along the solution with vanishing Mach
number at infinity gives $\calM (x)\sim \lambdat e^{-(\chi
  +\MR)/x}/x^2$.  Once the Mach number radial profile is known, the
density profile of the accreting gas is obtained from the analogous of
eq. (9), with
\begin{equation}
\rhotil(x)={\lambdat\over x^2\calM (x)}. 
\end{equation}

\section{The two-component galaxy models}

We now extend the results in Section 2.1, pertinent to isothermal
accretion in the one-component Jaffe model, to the family of
two-component JJ models presented in CZ18.  These models are
characterized by a {\it total} density distribution (stars plus dark
matter) $\rhog$ described by a Jaffe profile of total mass $\Mg$ and
scale-length $\rg$; the stellar density distribution $\rhos$ is also
described by a Jaffe profile of stellar mass $\Ms$, and scale radius
$\rs$.  The velocity dispersion anisotropy of the stellar component is
described by the Osipkov--Merritt formula (Merritt 1985), and a MBH is added at
the center of the galaxy.  Remarkably, almost all stellar
dynamical properties of JJ models with a central MBH can be expressed
by analytical functions.

The accretion solution of CP17 for a MBH at the center of a
Jaffe potential fully applies to JJ models, that are the first family
of two-component galaxy models with a central MBH where both the
fluidodynamics of accretion and the dynamics of the galaxy can be
described in an analytical way. In practice, for isothermal accretion
in the JJ models, there is the unique opportunity to easily compare the dynamical properties of
the stellar component of the galaxy (velocity dispersion, MBH sphere
of influence, etc.) with the accretion flow properties (the sonic
radius, the Bondi radius, the critical accretion parameter, the Mach
number profile, etc.).  We take here an important step further, and
fix the constant gas temperature 
$\Tinf$ by using the virial temperature of the stellar component.
In this way JJ models not only provide a more realistic
potential well for accretion, but also determine the accretion
temperature itself, yielding a natural ``closure'' for the problem
and fully constraining the solution. In order to link the properties
of JJ models to the general solution given in CP17, in the following
two Sections we introduce the properties of JJ models that are relevant
for the present study.

\subsection{Structure of the JJ models}

The density distribution of the stellar component of JJ galaxies is 
\begin{equation}
\rhos(r)={\Ms \rs\over 4\pi r^2(\rs + r)^2}=
{\rhon\over s^2(1+s)^2},\quad  s\equiv {r\over\rs},
\end{equation}
where $\Ms$ is the total stellar mass and $\rs$ is the scale-length;
the effective radius $\reff$ of the stellar profile is $\reff\simeq
0.7447\rs$.
We adopt $\Ms$ and $\rs$ as the mass and length scales, and define
\begin{equation}
\rhon\equiv{\Ms\over 4\pi\rs^3},\quad 
\Psin\equiv{G\Ms\over\rs},\quad 
\mu\equiv{\Mbh\over\Ms},
\end{equation} 
as the density and potential scales, and the last parameter
measures the MBH-to-galaxy stellar mass ratio. After introducing the
structural parameters (cfr. with eq. 12)
\begin{equation}
\MRg\equiv 
{\Mg\over\Ms},\quad \csig \equiv {\rg\over\rs}
\end{equation} 
we can give the {\it total} density distribution (stars plus dark
matter), that is 
also described by a Jaffe profile of scale lenght $\rg$ and total mass
$\Mg =\Ms+\MD$:
\begin{equation}
\rhog(r)= {\rhon\MRg\csig\over s^2(\csig+s)^2}. 
\end{equation}
From the request that the dark halo has a non-negative total
mass $\MD$, it follows that $\MRg\geq 1$ (see eq. 22). The cumulative mass
within the sphere of radius $r$, and the associated
gravitational potential, are given by
\begin{equation}
\Mg(r)={ \Ms\MRg s\over \csig+s},\quad 
\Phig (r)= {\Psin\MRg\over\csig}\ln {s\over \csig+s},
\end{equation}
and the analogous quantities for the stellar component are obtained
from eq. (24) for $\MRg=\csig=1$.  It also follows that the half-mass
(spatial) radius of the total mass profile is $\rg$, and it is $\rs$
for the stellar mass.

The density distribution $\rhoDM$ of the dark halo is given by
\begin{equation}
\rhoDM(r)={\rhon\over s^2}
\left[{\MRg\csig\over (\csig+s)^2} - {1\over (1+s)^2} \right],
\end{equation} 
so that $\rhoDM$ of JJ models {\it is not a Jaffe profile, unless the
  stellar and total length scales are equal} ($\csig=1$); the total
mass associated with $\rhoDM$ is $\MD =\Ms (\MRg -1)$. The request of
a non-negative $\MD$ does not prevent the possibility of an
unphysical, {\it locally negative} dark matter density, for an
arbitrary choice of $\MRg$ and $\csig$. In fact, CZ18 showed that the
condition to have $\rhoDM \ge 0$ at all radii is
\begin{equation}
\MRg \ge\Rm\equiv\max\left({1\over \csig},\csig \right).
\end{equation}
This condition implies also a monotonically decreasing $\rhoDM$, with
important dynamical consequences.  A dark halo of a model with
$\MRg=\Rm$ is called a {\it minimum halo} model. In the following we
use the parameter $\alpha$ to measure how much $\MRg$ is larger than
the minimum halo mass model, for assigned $\csig$:
\begin{equation}
\MRg=\alpha \Rm,\quad \alpha\geq 1.
\end{equation}
The special value $\csig=1$ corresponds to the minimum value $\Rm=1$,
i.e., to stellar and total densities that are proportional; in this
case, eq. (22) shows that for $\alpha=1$ there is no dark matter, and
one recovers the one-component Jaffe model used in CP17. The
properties of the dark halo profile as a function of $\csig$ and
$\MRg$ are fully discussed in CZ18. We stress that {\it everywhere in
  this paper we restrict to a dark profile more diffuse
  than the stellar mass}, which is obtained for $\csig\geq 1$; this choice
corresponds to the common expectation for real galaxies\footnote{The
  extension of the analysis to the cases $0\leq\csig <1$ would be
  immediate.}. From eqs. (26) and (27), then, in the following
we always have that $\MRg=\alpha\csig$.

It can be of interest for applications to evaluate the dark matter
fraction within a sphere of chosen radius. This amount is easily
calculated from eq. (24) as:
\begin{equation}
{\MD(r)\over\Mg(r)} = 1 - {\csig +s \over \alpha \csig (1+s)}, 
\end{equation}
where $\MD(r)=\Mg(r)-\Ms(r)$. Figure 1 shows the ratio between the
dark mass and the total mass within a sphere of radius $r=\reff$, as a
function of $\csig$, for various values of $\alpha$: the minimum halo
case ($\alpha =1$), and two cases (blue and red curves) with
$\MRg>\Rm$. Dark matter fractions below unity are easily obtained,
with low fractions ($<0.4$) for the minimum halo case. These low
values agree with those required by the modeling of the dynamical
properties of nearby early-type galaxies, that indicate the dark mass
within $\reff$ to be lower than the stellar mass (e.g., Cappellari et
al. 2015).

%%%%%%%%%%%%%%%%%%%%%%%%%%%%%%
\begin{figure}
\hskip -0.2truecm 
\includegraphics[height=0.55\textwidth, width=0.55\textwidth]{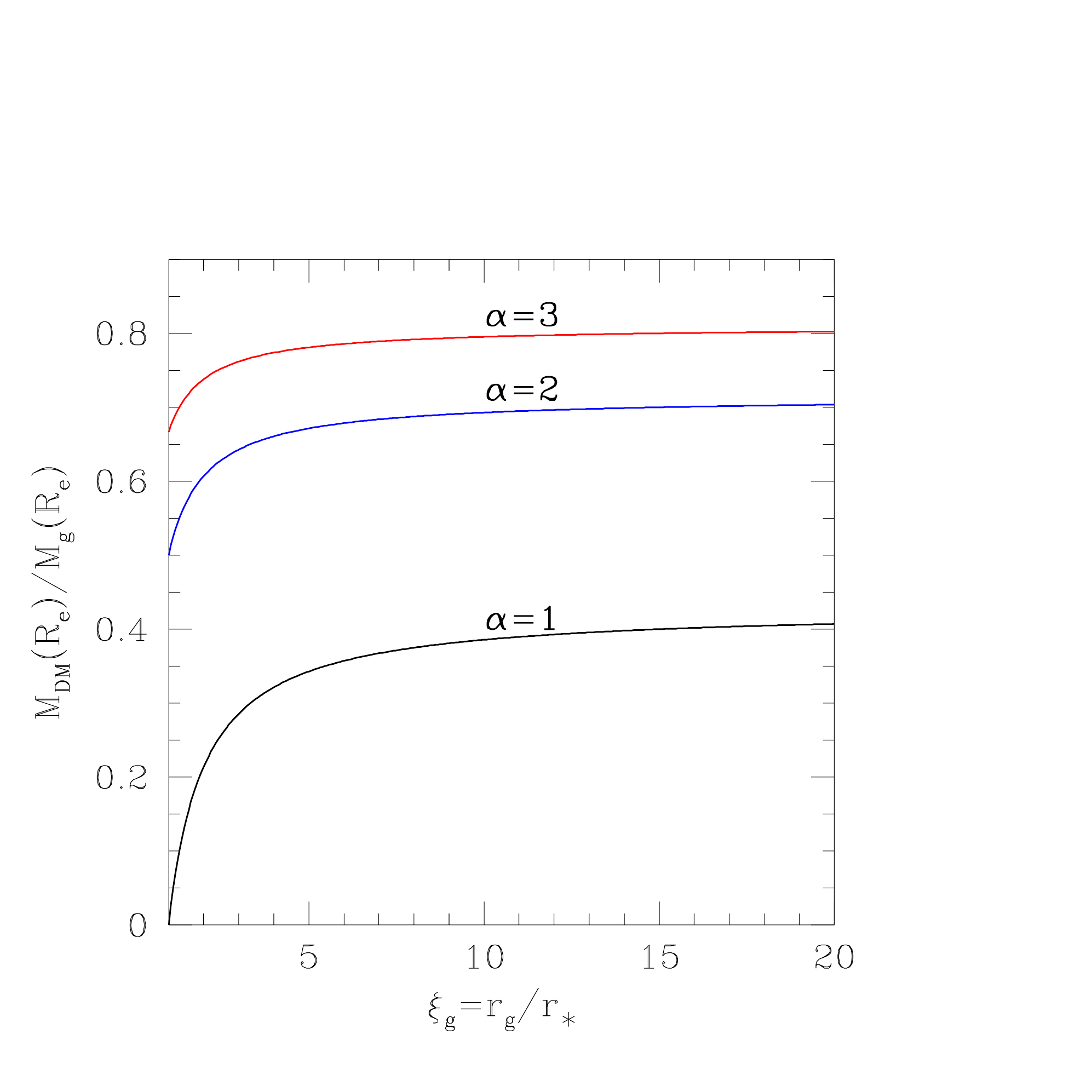}
\caption{Ratio between the dark mass and the total mass of JJ models
  (eq. 28), within a sphere of radius $r=\reff\simeq 0.75\rs$, as a
  function of $\csig$, for the minimum halo case ($\alpha =1$, black),
  and two non-minimum halo cases ($\alpha=2$, blue; and $\alpha=3$,
  red). The dark mass fraction is zero for the one-component (stellar)
  model, obtained for $\csig=1$ and $\alpha=1$.}
\label{}
\end{figure}
%%%%%%%%%%%%%%%%%%%%%%%%%%%%%%

For what concerns the radial profile of $\rhoDM$, for $r\to\infty$,
eq. (25) shows that $\rhoDM \sim (\MRg\csig -1)\rhos\propto r^{-4}$,
and so the densities of the dark matter and of the stars in the outer regions
are proportional.  For $r\to 0$, in non minimum-halo models $\rhoDM
\sim (\MRg/\csig-1)\rhos \propto r^{-2}$, while in the minimum-halo
models $\rhoDM \sim 2(1-1/\csig) s \rhos\propto r^{-1}$, so that these
latter models are centrally baryon-dominated, being $\rhos\propto
r^{-2}$.  It is interesting to compare the dark halo profile of JJ
models with the NFW profile, of total mass $\MD$, that we rewrite for
$r <\rt$ (the so-called truncation radius) as:
\begin{equation}
\rhoNFW(r)={\rhon(\MRg -1)\over f(c)s(\csih +s)^2},\quad f(c)=\ln(1+c) -{c\over 1+c},
\end{equation}
where $\csih \equiv r_{\rm NFW}/\rs$ is the NFW scale-length $r_{\rm
  NFW}$ in units of $\rs$, and $c\equiv\rt/r_{\rm NFW}$. From the
considerations above about the behavior of $\rhoDM(r)$ at small and
large radii, it follows that $\rhoDM$ and $\rhoNFW$ at small and large
radii cannot in general be similar. However, in the case of the
minimum-halo, near the center $\rhoDM\propto r^{-1}$, and it can be
proven that $\rhoDM$ and $\rhoNFW$ can be made identical for $r\to 0$
by fixing
\begin{equation}
\csih =\sqrt{{\csig\over 2f(c)}}
\end{equation}
in eq. (29).  Therefore, once a specific JJ minimum halo model is
considered, eqs. (29)-(30) allow to determine the NFW profile that has
the same total mass and central density profile as $\rhoDM$. It turns
out that the dark halo profiles of JJ minimum halo models are
surprisingly well approximated by the NFW profile over a very large
radial range, for realistic values of $\csih $ and $c$ (CZ18).

\subsection{Dynamics of the JJ models}

CZ18 present and discuss the analytical solutions of the Jeans
equations and all the dynamical properties of Osipkov-Merritt
anisotropic JJ models with a central
MBH of mass $\Mbh$, where the total gravitational potential is
\begin{equation}
\Phi_{\rm T} (r)=\Phig (r)- {\Psin \mu\over s}. 
\end{equation}
The radial component of the {\it stellar} velocity dispersion tensor
is given by $\srad^2 (r)=\sigg^2 (r)+\sigBH^2(r)$, 
where $\sigg$ indicates the contribution due to
$\Phig (r)$, and $\sigBH$ the contribution due to
the MBH.  Under the assumption of Osipkov-Merritt anisotropy,
$\sigg  (0)$ is coincident with that of the isotropic case
(except for purely radial orbits), independently of the anisotropy radius $\ra >0$, and is given by:
\begin{equation}
\sigg^2 (0)= {\Psin\MRg\over 2\csig}={\Psin\alpha\over 2},
\end{equation}
where in the last identity we use eq. (27), and restrict to
$\csig\geq 1$. Note that the value of $\sigg (0)$ is therefore
{\it independent} of $\csig$, for $\csig\geq 1$, and, in the minimum
halo model, it is {\it coincident} with that of the one-component
(purely stellar) Jaffe model.  The leading term of the MBH
contribution to $\srad$ near the center (except for the case of purely
radial orbits, $\ra =0$) coincides with the isotropic case
independently of the Osipkov-Merritt anisotropy radius, with
\begin{equation}
\sigBH^2 (r) \sim {\Psin\mu\over 3 s};
\end{equation}
at variance with $\sigg (r)$, it diverges as $\mu /r$ for $r\to
0$. Therefore, in presence of the central MBH, $\srad$ is dominated by
its contribution, and similarly is the projected velocity dispersion
($\siggp$).

In order to relate the models with observed quantities, it is helpful
to consider the {\it projected} velocity dispersion in the central
regions.  CZ18 shows that for $\ra >0$
\begin{equation}
\siggp (0)=\sigg (0)
\end{equation}
while,  independently of the value of $\ra \geq 0$
\begin{equation}
\sigBHp^2 (R)\sim {2\Psin\mu\over 3\pi}{\rs\over R},
\end{equation}
where $R$ is the radius in the projection plane.
The two equations above determine a fiducial value for the radius
($\Rinf$) of the so-called {\it sphere of influence}.  We define
operationally $\Rinf$ as the distance from the center in the
projection plane where the (galaxy plus MBH) projected velocity
dispersion $\sigp (R)$ equals a chosen fraction
$\epsilon$ of the projected velocity dispersion of the galaxy in
absence of the MBH. In practice, as $\Rinf<<\rs$, and in JJ models
without MBH the velocity dispersion profile flattens to a constant
value, we define
\begin{equation}
\sigp (R)\simeq \sqrt{\sigBHp^2 (\Rinf)+\siggp^2 (0)}=(1+\epsilon)\siggp(0),
\end{equation}
and from eqs. (34)-(35) one has:
\begin{equation}
{\Rinf\over\rs}=
{4\csig\mu\over 3\pi\MRg\epsilon(2+\epsilon)}=
{4\mu\over 3\pi\alpha\epsilon(2+\epsilon)},
\end{equation}
where the last identity holds for $\csig\geq 1$. For realistic values
of the parameters, $\Rinf$ is of the order of a few pc (see Sect. 5.1
for a more quantitative discussion).

As anticipated in the Introduction, a reasonable estimate of the gas
temperature, supported by observations (e.g., Pellegrini 2011), is
given by the stellar virial temperature $\Tv=<\mu>m_p\sigv^2/3$. 
The definition of $\sigv$ comes from the virial theorem that, for
the stellar component, reads:
\begin{equation}
2\Ks =-\Wg-\Wbh,
\end{equation}
where $\Ks\equiv \Ms\sigv^2/2$ is the total kinetic energy of the stars, and
\begin{equation}
\Wg= - 4\pi G\int_0^{\infty} r\rhos (r)\Mg(r)dr
\end{equation}
is the interaction energy of the stars with the total gravitational field of
the galaxy (stars plus dark matter), and finally
\begin{equation}
\Wbh=-4\pi G \Mbh \int_0^\infty r \rhos (r) dr
\end{equation}
is the interaction energy of the stars with the central MBH. Note that
$\Wbh$ diverges, because the stellar density profile diverges near the
origin as $ r^{-2}$; instead, $\Wbh$ converges for $\gamma$ models
with $0\le \gamma<2$.  Since we will use $\Ks$ to evaluate the gas
temperature over the whole body of the galaxy (where the MBH effect is
negligible), we only consider $\Wg$ in the determination of
$\sigv$. Therefore $\sigv^2\equiv -\Wg/\Ms$, where from CZ18:
\begin{equation}
\Wg = -\Psin\Ms\MRg\Wgt,\quad
\Wgt ={\csig-1-\ln \csig\over (\csig-1)^2},
\end{equation}
and $\Wgt (1) =1/2$. It follows that $\sigv^2=\Psin\MRg\Wgt
=\Psin\alpha\calFg(\csig)$, where we introduced the function
$\calFg(\csig)\equiv\csig\Wgt$.  Note that $\calFg(\csig)$ increases
from $\calFg(1)=1/2$ to $\calFg(\infty)=1$. In practice, at fixed
$\alpha$ and increasing $\csig$, eq. (27) dictates that $\MRg$
increases to arbitrarily large values, but since $\calFg \to 1$, then
$\Wg$ and $\sigv$ (and so the mass-weighted squared escape velocity)
remain limited. Physically, this is due to the fact that more massive
halos are necessarily more and more extended, due to the request for
positivity in eq. (26), with a compensating effect on the depth of the
total potential. Moreover, from eqs. (32) and (34) it follows that
$\sigv^2=2\calFg(\csig)\siggp^2(0)$, so that in JJ models without MBH,
$\sigv$ is just proportional to the stellar central projected velocity
dispersion, and the proportionality constant is a function of $\csig$
{\it only}, with $\sigv =\siggp (0)$ for $\csig=1$.

%%%%%%%%%%%%%%%%%%%%%%%%%%%%%%%%%%%%%%%%%%%%%%%%%%%%%%%%%%%%%%%%%%%%%%
\renewcommand\arraystretch{1.4}
\begin{table}
\centering 
\caption{List of parameters}
\vspace{2mm}
\begin{tabular}{cc}
\toprule 
 \midrule                                                                                                                                
Symbol      &  Description  \\ 
 \midrule                                                                                                                                
 \midrule                                                                                                                                
Galaxy structure: & \\
 \midrule                                                                                                                                
$\Mg$   &   total mass         \\
$\Ms$   &  stellar  mass         \\                                                                                
$\Mbh$   &  central MBH mass         \\                                                                                
$\rg$   &    total density scale-length            \\
$\rs$   &   stellar density scale-length            \\
$\sigv$   &   stellar virial velocity dispersion           \\
$\Tv$   &   stellar virial temperature          \\
\midrule                                      
$\mu$   &   $\Mbh/\Ms$         \\                  
$\csig$   &   $\rg/\rs$           \\
$\MRg$   &   $\Mg/\Ms$ \\
 & \quad ($=\alpha\csig$,\quad $\csig\geq 1$,\quad $\alpha\geq 1$)         \\
\midrule\midrule                  
Accretion flow: & \\
\midrule                                                                                                                
$\Tinf$   &   temperature ($=\beta\Tv$,\quad $\beta >0$)        \\
$\csinf$   &  sound velocity        \\                                                                                
$\rb$   &  Bondi radius        \\                                                                                
$\rmin$   & sonic radius           \\
$\Mdott$   & mass accretion rate          \\
\midrule
$\calM$   & Mach number          \\
$\lambdat$   & critical accretion parameter           \\
$\MR$   &  $\Mg/\Mbh$        \\
$\xi$   &   $\rg/\rb$         \\
$\xmin$   &   $\rmin/\rb$         \\
$\betac$ & critical $\beta$ ($=3/(2\calFg$)) \\
\bottomrule 
\end{tabular}
\flushleft 
%Notes: $(1)$ bla bla bla 
\label{tab:params}
\end{table}
\renewcommand\arraystretch{1.}
%%%%%%%%%%%%%%%%%%%%%%%%%%%%%%%%%%%%%%%%%%%%%%%%%%%%%%%%%%%

\section{Linking stellar dynamics to fluidodynamics}

In the general solution of CP17, once that the parameters $\cal R$ and
$\xi$ in eq. (12) are assigned, and the Jaffe structural scales $\Mg$ and
$\rg$ characterizing the {\it total} galactic potential are chosen,
the gas temperature $\Tinf$ remains fixed, because $\xi
=\rg/\rb$. Therefore, generic values of $\xi$ can easily correspond to
unrealistic values of the gas temperature.  Clearly, JJ models offer
an interesting possibility: as the dynamical properties of the stellar
component of the galaxy can be analytically calculated once the total
potential (due to stars, dark matter and central MBH) is assigned, the
Virial Theorem for the stellar component can be used to compute the
virial ``temperature'' $\Tv$ of the stars, a realistic proxy for the gas
temperature $\Tinf$; then, the CP17 solution for the accreting gas in the 
total Jaffe potential of given $\Tv$ can be built. In
practice, the idea is to self-consistently ``close'' the model,
determining a fiducial value for the gas temperature as a function of
the galaxy model hosting accretion. In this approach, the steps to
build an accretion solution are: 1) choose $\Ms$, $\rs$,
$\mu$, $\MRg$ and $\csig$ for a realistic galactic model; 2) obtain
the gas virial temperature $\Tv$; 3) derive $\MR$ and
$\xi$ to be used in the Bondi problem, and construct the corresponding
CP17 solution.

Suppose the galaxy parameters in the first step are given. Then, for
assigned $\MRg$, $\csig$ and $\mu$, the parameter $\MR$ in the
accretion solution is obtained from eqs. (12) and (21)-(22) as
\begin{equation}
\MR={\MRg\over\mu} = {\alpha\csig\over\mu},
\end{equation}
where the last expression depends on the fact that we restrict to the
case $\csig\geq 1$. Since $\mu\approx 10^{-3}$, and $\MRg$ is expected (say)
in the range $1\div 20$, the $\MR$ values fall in the range $10^3\div
10^4$ (see also Sect. 5.1 for a more quantitative discussion).

%%%%%%%%%%%%%%%%%%%%%%%%%%%%%%
\begin{figure*}
\hskip -0.2truecm 
\includegraphics[height=0.5\textwidth, width=0.5\textwidth]{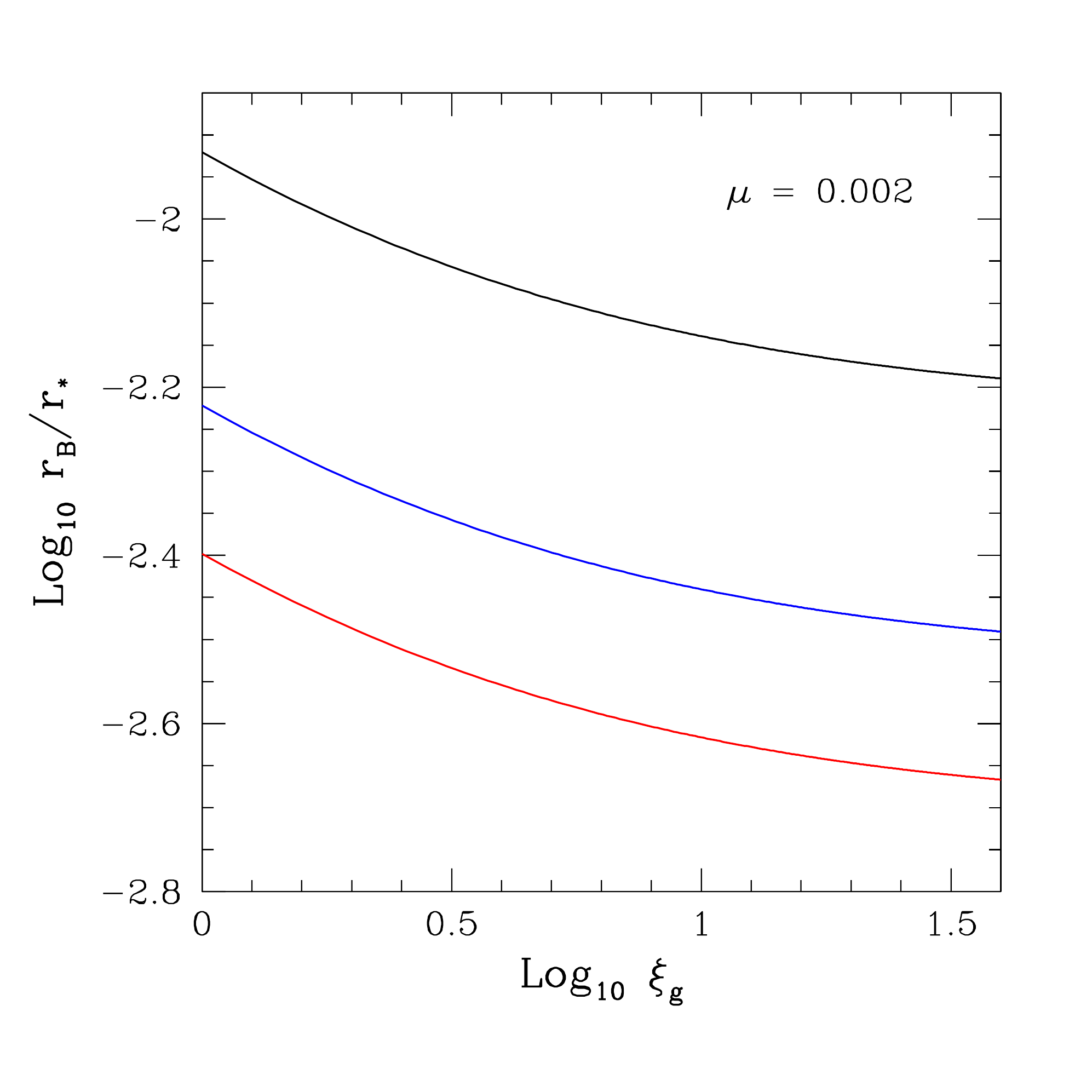}
\includegraphics[height=0.5\textwidth, width=0.5\textwidth]{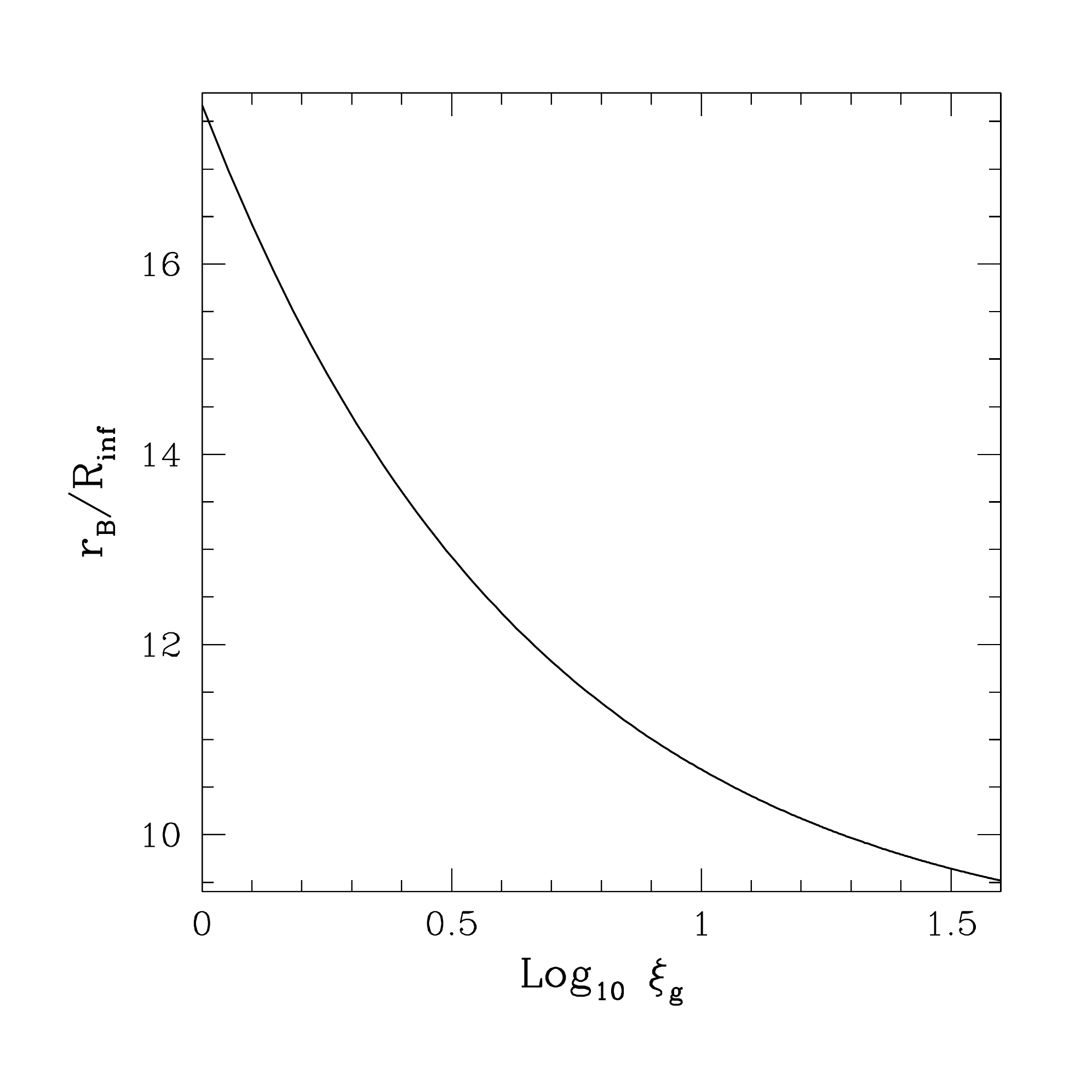}
\includegraphics[height=0.5\textwidth, width=0.5\textwidth]{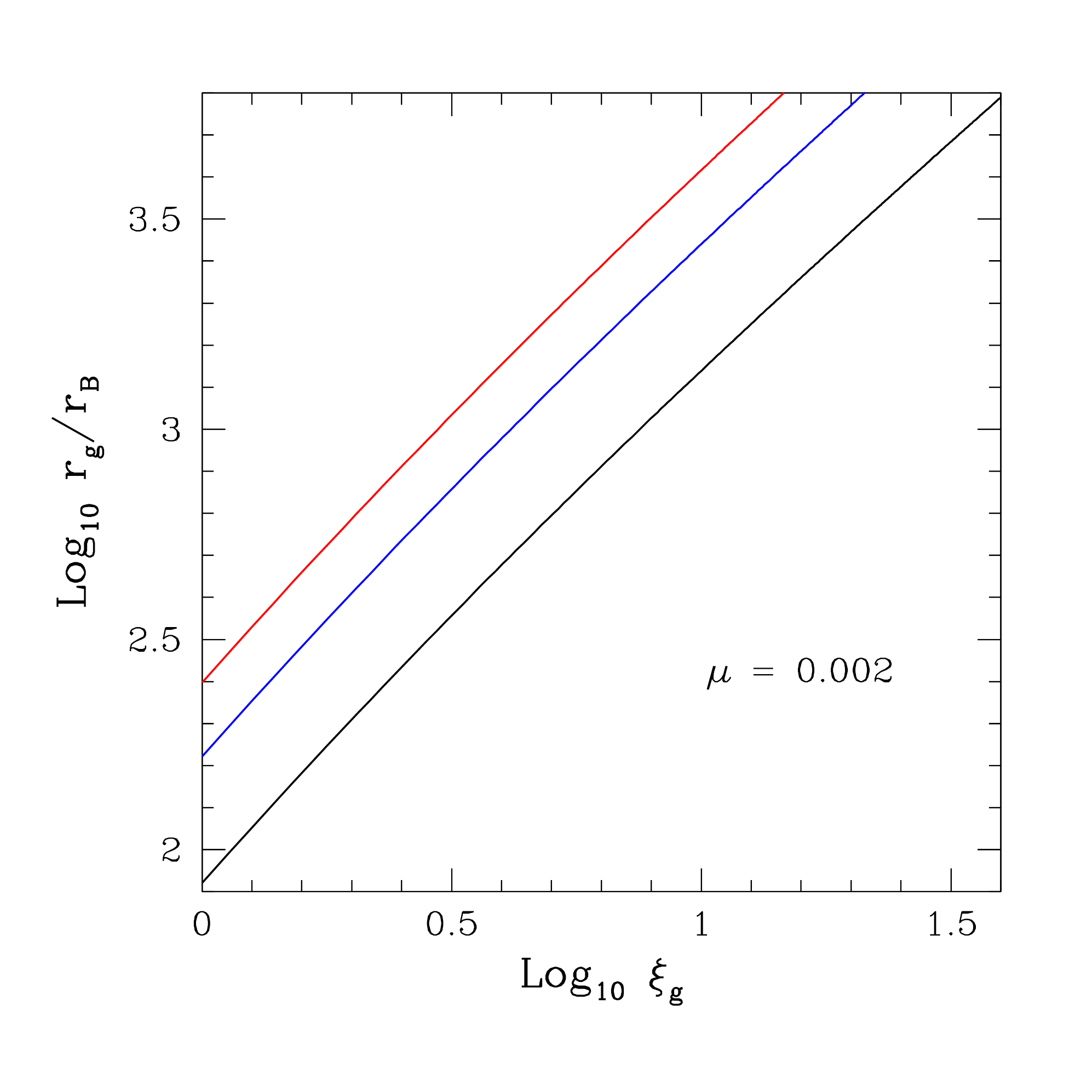}
\includegraphics[height=0.5\textwidth, width=0.5\textwidth]{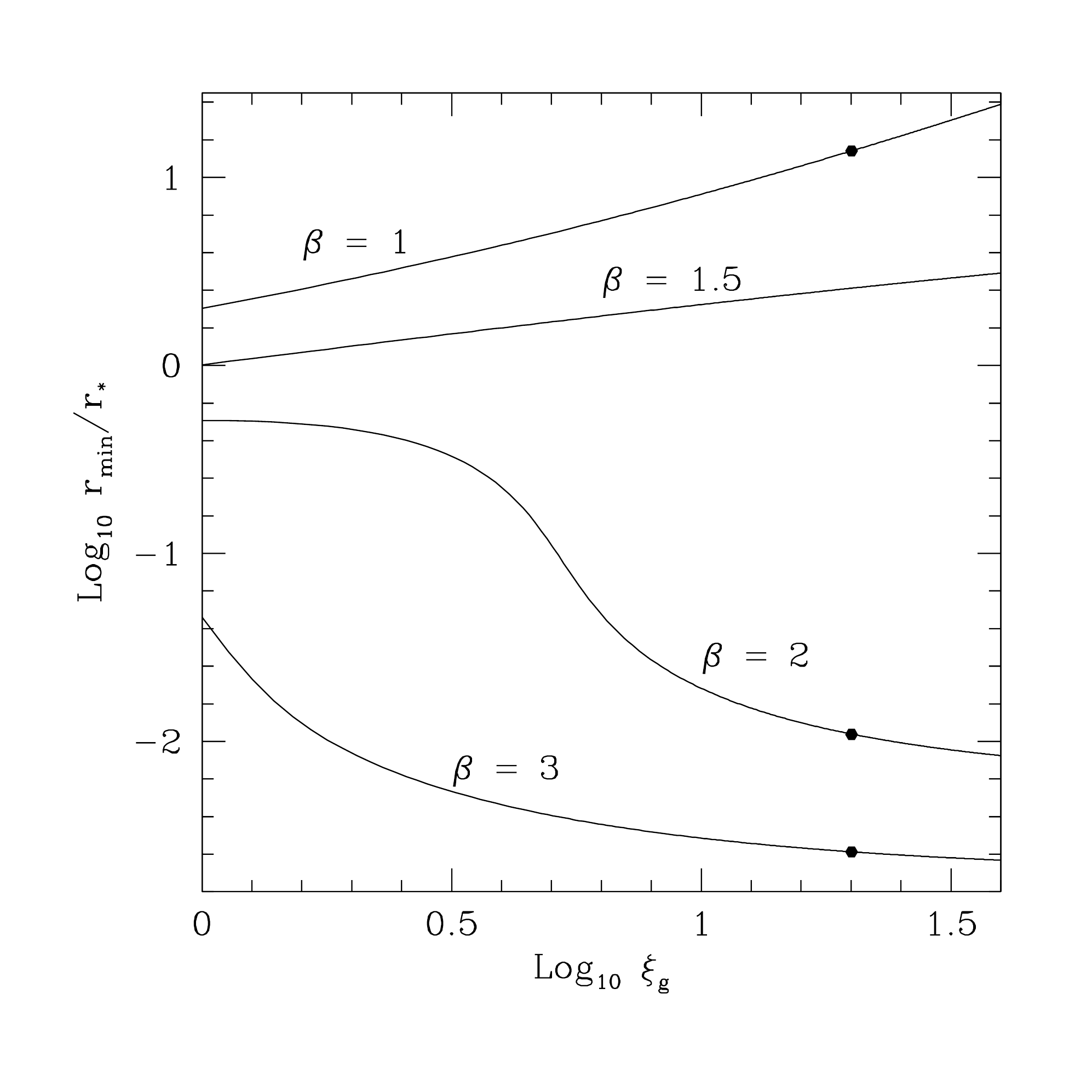}
\caption{Relevant scale-lengths of the isothermal accretion solution
  in JJ models, as a function of $\csig=\rg/\rs$. A MBH-to-galaxy
  stellar mass ratio $\mu=2\times 10^{-3}$ is assumed, and
  $\Tinf=\beta\Tv$.  Top left: the ratio $\rb/\rs$ from eq. (44) with
  $\beta=1$ and $\alpha =1,2,3$, following the color scheme as in
  Figure 1.  Top right: ratio $\rb/\Rinf$ of the Bondi radius to the
  radius of the sphere of influence of the MBH, $\Rinf$, for $\epsilon
  = 0.5$ from eqs. (37) and (44), with $\beta =1$. Note that this
  ratio is independent of $\alpha$ and $\mu$.  Bottom left: the
  parameter $\xi=\rg/\rb$ from eq. (46) with $\beta =1$ and $\alpha
  =1,2,3$.  Bottom right: ratio $\rmin/\rs$ for $\alpha =1$ and $\beta
  =1,3/2,2,3$: for large values of $\MR$ and $\beta <\betac$ the ratio
  is almost independent of $\alpha$ but strongly dependent on gas
  temperature, as follows from eqs. (44), (47) and (48). Solid dots
  ($\rmin/\rs \simeq 13.84, 0.011, 0.0026$) mark the position of the
  sonic radius for the minimum halo case with $\csig =20$.}
\label{}
\end{figure*}
%%%%%%%%%%%%%%%%%%%%%%%%%%%%%%

From eq. (12), the second parameter ($\xi$) characterizing the
accretion solution requires the computation of the Bondi radius $\rb$,
and then of the gas temperature $\Tinf$; we choose 
$\Tinf=\beta\Tv$, with $\beta >0$ a dimensionless parameter. Thus the
isothermal sound speed is given by:
\begin{equation}
\csinf^2={k \Tinf\over <\mu> \mpr}={\beta\sigv^2\over 3},
\end{equation}
and from eqs. (3) and (41) the Bondi radius reads:
\begin{equation}
{\rb\over\rs}={3\mu\over \alpha\beta\calFg},\quad
\calFg\equiv\csig\Wgt (\csig).
\end{equation}
From the behavior of the function $\calFg(\csig)$, it follows that at
fixed $\alpha$ and $\beta$ one has:
\begin{equation}
{3\mu\over\alpha\beta}\leq{\rb\over\rs}\leq {6\mu\over\alpha\beta},
\end{equation}
where the lower limit is obtained for $\csig\to\infty$, and the upper
limit for $\csig=1$; in this latter case, $\alpha =1$ gives the value
of $\rb/\rs$ in a one-component (stellar) Jaffe galaxy.  As expected,
$\rb$ decreases for increasing $\alpha$, $\beta$ and $\csig$, i.e.,
for increasing $\Tinf$. Figure 2 (top left panel) shows the trend of
$\rb/\rs$ with $\csig$, in the minimum halo case ($\alpha =1$), and
for $\alpha=2$ and $3$, when $\beta=1$ and $\mu=0.002$. Therefore, in
a real galaxy with $\rs$ of the order of a few kpc, and with a gas
temperature of the order of $\Tv$, $\rb$ is of the order of tens of pc
(see also Sect. 5.1 for a more quantitative discussion).  As an
additional information on $\rb$, in Fig. 2 (top right panel) we show
the trend of $\rb/\Rinf$ with $\csig$, for $\beta =1$; note that from
eqs. (37) and (44), the ratio is independent of $\alpha$ and $\mu$, so
that only one curve is plotted; higher values of $\beta$ correspond to
smaller values of $\rb$.

Using eqs. (44) and (12), we finally obtain the expression
\begin{equation}
\xi ={\alpha\beta\csig\calFg\over 3\mu}={\MR\beta\calFg\over 3}.
\end{equation}
It follows that $\xi \to \alpha\beta/(6\mu)$ for $\csig\to 1$, while
it grows without bound for $\csig\to\infty$, as
$\xi\sim \alpha\beta\csig /(3\mu)$.  In practice, at variance with the
general cases in KCP16 and CP17, now $\MR$ and $\xi$ are linked, and
increasing values of $\MR$ correspond to increasing values of
$\xi$. The list of all parameters introduced in this work is given in
Table 1.  Figure 2 (bottom left panel) shows the trend of $\xi $ with
$\csig$, in the minimum halo case ($\alpha =1$), and for $\alpha=2$
and $3$, and $\beta =1$; $\rb$ is of the order of $10^{-3}\rg$; higher
values of $\beta$ correspond to larger values of $\xi$.

Having obtained the expressions for $\MR$ and $\xi$ as a function of
the model parameters, a few considerations are in order.  The first is
that in JJ models isothermal accretion is {\it always} possible in
absence of a central MBH and $\beta =1$, because the accretion
condition in eq. (17) is automatically satisfied by the virial
temperature of the stellar component, when $\Tinf=\Tv$. By allowing
for a $\Tinf >\Tv$, it is easy to show that Bondi isothermal accretion
in {\it absence} of a central MBH (or when $\chi =0$) is possible in
JJ models only for gas temperatures lower than a critical value,
i.e. only for $\beta \leq \betac\equiv 3/(2\calFg)$. From the behavior
of $\calFg$ it follows that $3/2\leq\betac\leq 3$, where the lower
limit corresponds to $\csig\to\infty$ and the upper limit to
$\csig =1$.  The second is that from eq. (46) the ratio
$\MR/\xi =2\betac/\beta$ appearing in the definition of $f(x)$ in
eq. (13), depends on $\beta$ and $\csig$ {\it only}, and it shows that
for very large values of $\beta$ the problem reduces to the classical
Bondi accretion.

The critical value $\betac$ plays an important role also in models
{\it with} a central MBH, determining a particular temperature at
which there is a sudden transition in the position of the sonic
radius. In fact, the location of $\rmin$ in terms of the scale-length
$\rs$, is given by
\begin{equation}
{\rmin\over\rs}=\xmin(\chi,\MR,\xi) \, {\rb\over\rs},
\end{equation}
where $\xmin$ is given by eq. (15), and can be easily computed
analytically once $\MR$ and $\xi$ are determined.  Figure 2 (bottom
right panel) shows $\rmin/\rs$ as a function of $\csig$, for
$\alpha =1$ and different values of $\beta$. The most relevant feature
is the considerable variation in the position of $\rmin$, from very
external to very inner regions, for $\beta$ increasing. Equation (47)
shows that $\rmin$ is determined by the combined behavior of two
functions, namely $\xmin$ and $\rb/\rs$; we now focus on $\xmin$,
having already established that $\rb\propto 1/\beta$, and thus the
large variation of $\rmin$ can only in part be due to the dependence
of $\rb$ on $\beta$. Instead, from eqs. (15), (42) and (46), for
$\MR\to\infty$ and fixed\footnote{From eq. (46)
  $2\xi=\MR\beta/\betac$, and from eq. (15) it follows immediately
  that the limit for $\MR\to\infty$ is not uniform in $\beta$.}
$\beta$, $\xmin$ is given by:
\begin{equation}
\xmin\sim 
\begin{cases}
    \displaystyle{{\MR\; (1- \beta/\betac)\over 2},\quad \beta<\betac,}\\
   \displaystyle{ {\sqrt{\chi\MR}\over 2},\quad \beta = \betac,}\\
  \displaystyle{{\chi \over 2(1 -\betac/\beta)},\quad \beta >\betac .}
\end{cases}
\end{equation}
Note that the limit $\MR\to\infty$ describes models with increasing
$\alpha$ at fixed $\mu$ and $\csig$, or with increasing $\csig$ at
fixed $\alpha$ and $\mu$, or a vanishing MBH mass at fixed $\alpha$
and $\csig$. As in the present models $\alpha\geq 1$, $\csig\geq 1$
and $\mu=0.002$, then $\MR$ is quite large, and the asymptotic trends
in eq. (48) already provide a reasonable approximation of the true
behavior, that is increasingly better for large values of $\alpha$ and
$\csig$, and small values of $\mu$.  Of course, an independent check
of the first two expressions above can be obtained by recovering them
from the exact eq. (16) (pertinent to a Jaffe galaxy with $\Mbh =0$),
by using eqs. (42) and (46) for vanishing MBH mass, i.e., $\mu\to 0$
and $\MR\to\infty$, and $\beta\leq\betac$.

Qualitatively, eq. (48) shows that for $\beta <\betac$, $\xmin$
increases as $\MR$. Instead, for $\beta >\betac$, $\xmin$ is
independent of $\MR$, and for very large values of the gas temperature
tends to $\chi/2$, the limit value of classical Bondi accretion with
electron scattering (e.g., CP17, eq. 25).  As Fig. 3 shows, even a
moderate increase in the gas temperature produces a sudden decrease in
the value of $\xmin$.  A numerical investigation of polytropic
accretion in JJ models shows that $\xmin$ suddenly drops to values
$\la 1$, as $\gamma$ increases with respect to the isothermal
case. This behavior is reminiscent of the sudden transition of $\xmin$
from external to internal regions in Hernquist galaxies, discussed in
CP17; in this case the transition is due to the existence, for
$\gamma >1$, of {\it two} minima for the polytropic function $f(x)$ of
the Jaffe potential (as obtained inserting eq. 11 in eq. 47 in
KCP16). In the polytropic Hernquist case, the two minima can be
present also in the isothermal case (CP17, Appendix B.2), while for
the Jaffe potential in the isothermal case there is only one minimum,
given in eq. (15).

It is now easy to explain the behavior of the curves in Fig. 2 (bottom
right panel), where several cases of eq. (47) are plotted.  For
example, from the first identity in eq. (48) and eqs. (42)-(44),
eq. (47) predicts that for $\MR\to\infty$ and $\beta<\betac$, we have
$\rmin/\rs\sim\csig (\betac/\beta -1)$, so that for $\csig\to \infty$
and $\beta = 1$, it results $\rmin/\rs\sim\csig/2$, while for
$\csig =1$ and large values of $\alpha/\mu$, one has
$\rmin\approx 2\rs$. For $\Tinf$ corresponding to the range
$3/2\leq\beta\leq 3$, there is a transition value of $\csig$ such
that, for larger $\csig$, $\betac$ drops below the adopted $\beta$,
and the third expression in eq. (48) applies. In these cases the sonic
radius moves to the central regions, with
$\rmin/\rs\sim \chi\mu/[\alpha (\beta/\betac -1)]$.

%%%%%%%%%%%%%%%%%%%%%%%%%%%%%%
\begin{figure}
\hskip -1truecm 
\includegraphics[height=0.55\textwidth, width=0.55\textwidth]{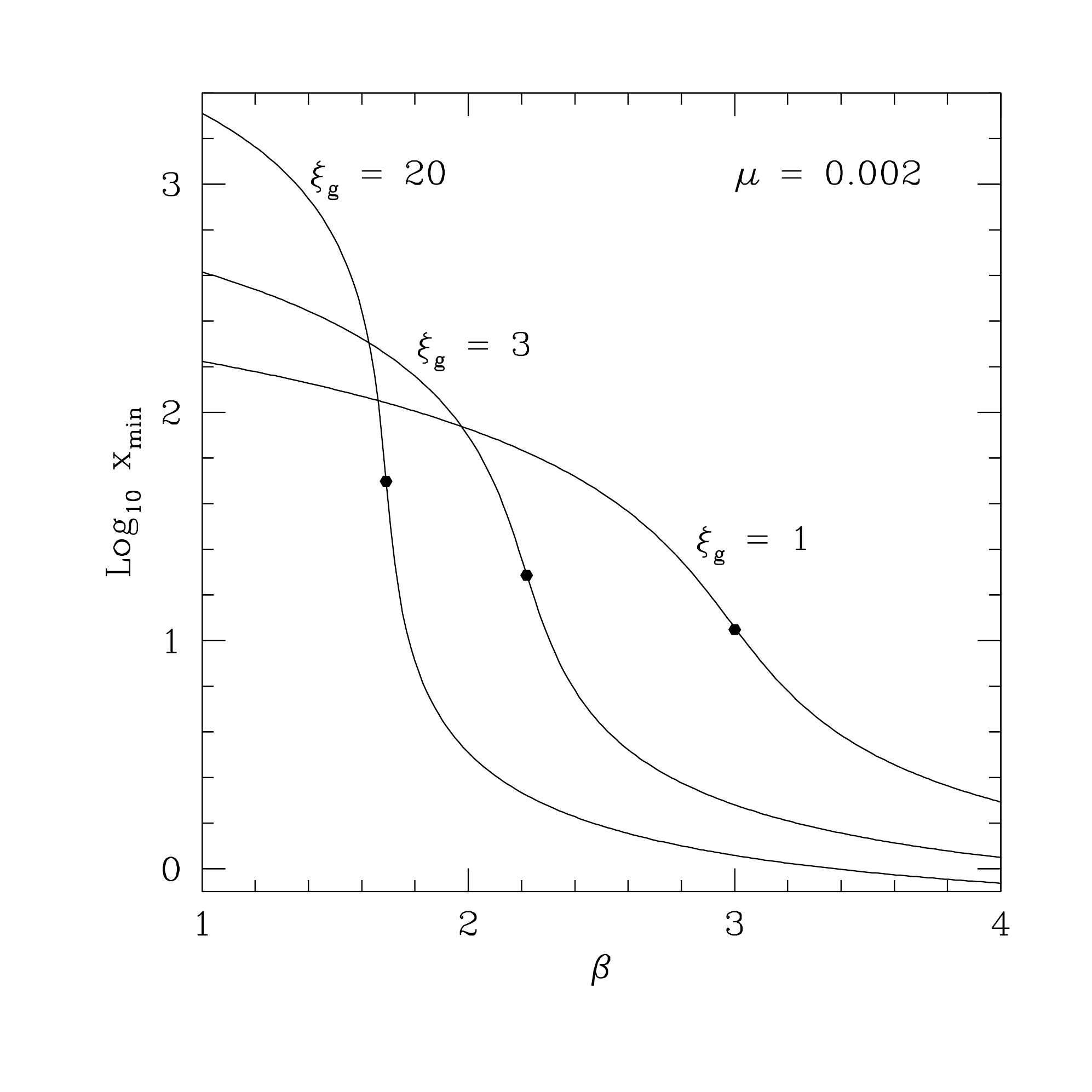}\\
\caption{Position of $\xmin$ as a function of the temperature
  parameter $\beta$, for $\mu=0.002$ and for three minimum halo models
  ($\alpha =1$), with $\csig =1,3,20$; these correspond to critical
  values of the temperature given by $\betac \simeq 3, 2.2, 1.7$,
  respectively (solid dots). The values of $\xmin$ are reproduced
  remarkably well by the asymptotic expressions in eq. (48).}
\label{}
\end{figure}
%%%%%%%%%%%%%%%%%%%%%%%%%%%%%%

Finally, Fig. 4 shows the trend of $\lambdat$ as a function of $\csig$,
for $\alpha =1,2,3$ and, when $\alpha =1$, for different 
gas temperatures as determined by the $\beta$ value (dotted
lines). In analogy with eq. (48), an asymptotic analysis shows that in
the limit of $\MR\to\infty$,
\begin{equation}
\lambdat\sim 
\begin{cases}
    \displaystyle{{\MR^2\;(1- \beta/\betac)^{2-{2\betac\over \beta}}\over 4\sqrt{e}}, 
\quad \beta<\betac,}\\
   \displaystyle{{\MR^2\over 4\sqrt{e}},\quad \beta = \betac,}
\end{cases}
\end{equation}
and for simplicity we do not report the expression of $\lambdat$ for
$\beta>\betac$, that however can be easily calculated. For fixed $\MR$
and $\chi$, very large $\beta$ correspond to
$\lambdat\sim\chi^2\lambdacr$, in accordance with the classical case
(KCP16, CP17). Equation (49) nicely explains the values and the trend
of $\lambdat$ with $\csig$ and $\alpha$, in particular the almost
perfect proportionality of $\lambdat$ to $\alpha^2\csig^2$. From
eq. (14), this implies that, for the same boundary conditions, the
true accretion rate $\Mdott$ for increasing  $\csig$ becomes
much larger than $\Mdotb$, the rate in sole presence of the MBH.

%%%%%%%%%%%%%%%%%%%%%%%%%%%%%%
\begin{figure}
\hskip -1truecm 
\includegraphics[height=0.55\textwidth, width=0.55\textwidth]{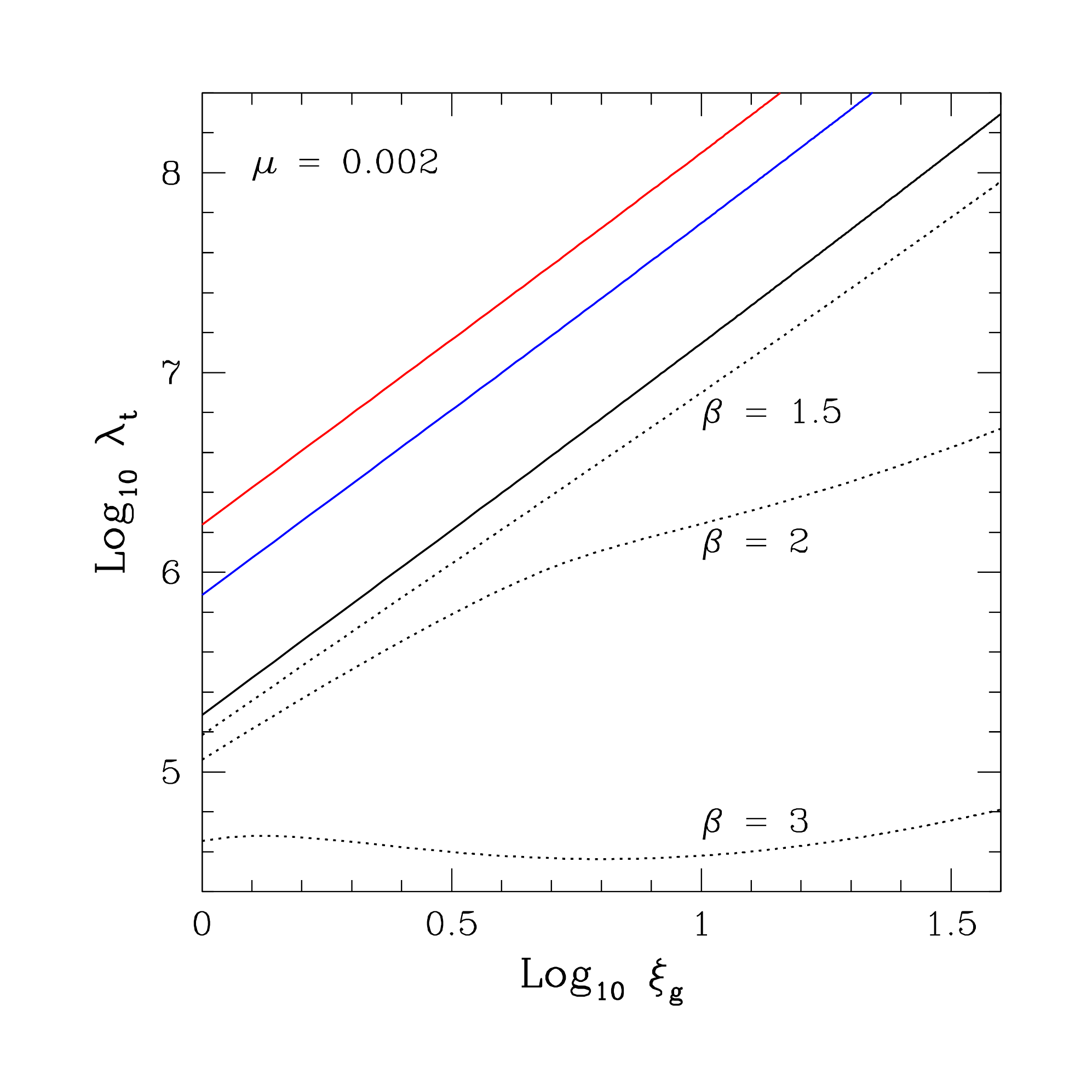}\\
\caption{The critical accretion parameter $\lambdat$ as a function of 
  $\csig$, for the miminum halo case $\alpha=1$ (black), and $\alpha =
  2,3$ (blue and red), and $\chi =1$ and 
  $\beta=1$. The dotted curves refer to $\alpha =1$ and three 
  different values of $\beta$.}
\label{}
\end{figure}
%%%%%%%%%%%%%%%%%%%%%%%%%%%%%%

\section{Two applications}

We present here two applications of the results above.  The first is a
practical illustration of how to determine the main parameters
describing the galactic structure, and the gas accretion in it, for JJ
models (see Table 1).  One will see how very reasonable values for the
main structural properties can be obtained, and then realistic galaxy
models can be built.  The second application considers the deviation
from the true value of the mass accretion derived using the density
along the accretion solution in JJ galaxies, and the framework of
classical Bondi theory.

%%%%%%%%%%%%%%%%%%%%%%%%%%%%%%
\begin{figure*}
\vskip -2truecm
\includegraphics[height=0.55\textwidth, width=0.55\textwidth]{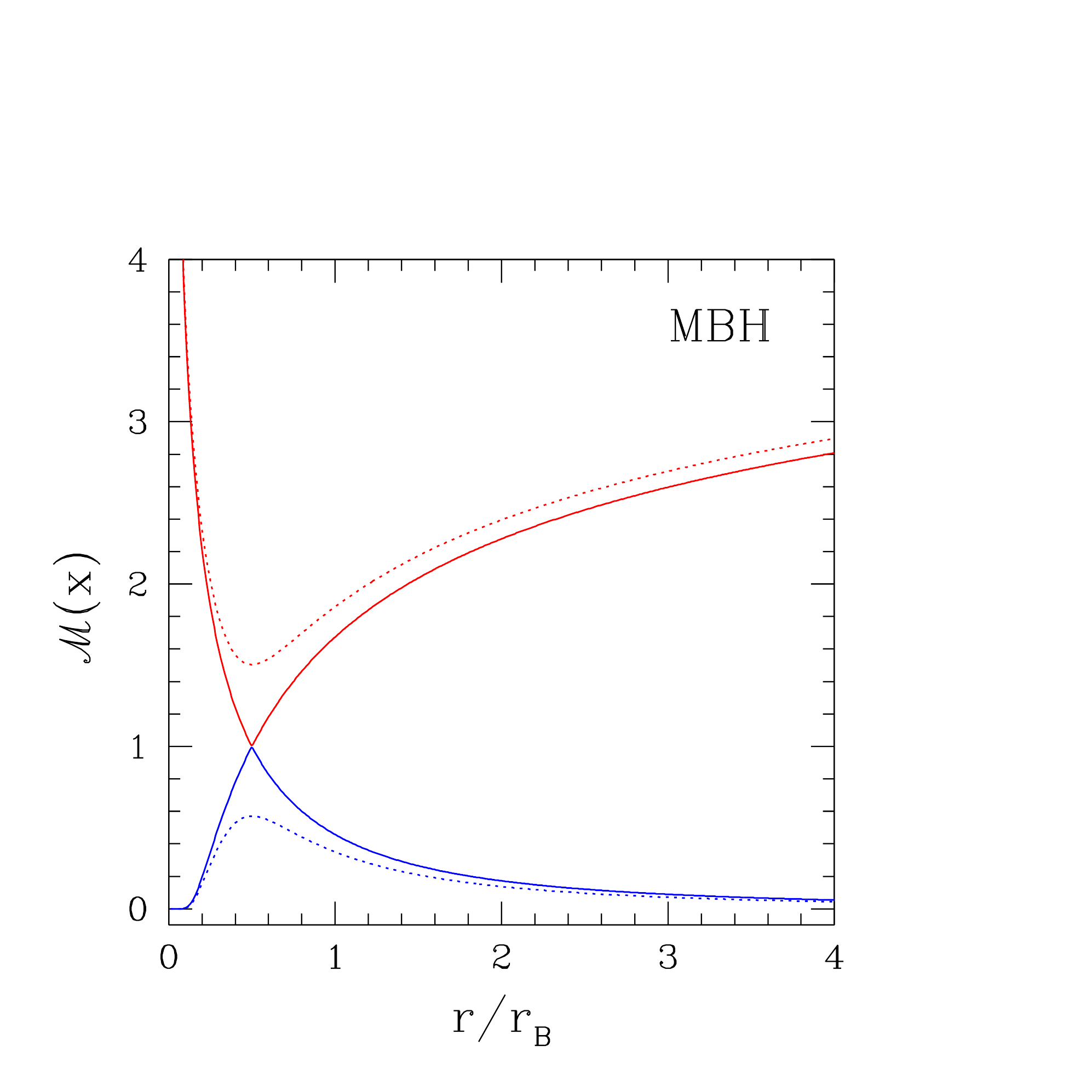}
\hskip -1truecm
\includegraphics[height=0.55\textwidth, width=0.55\textwidth]{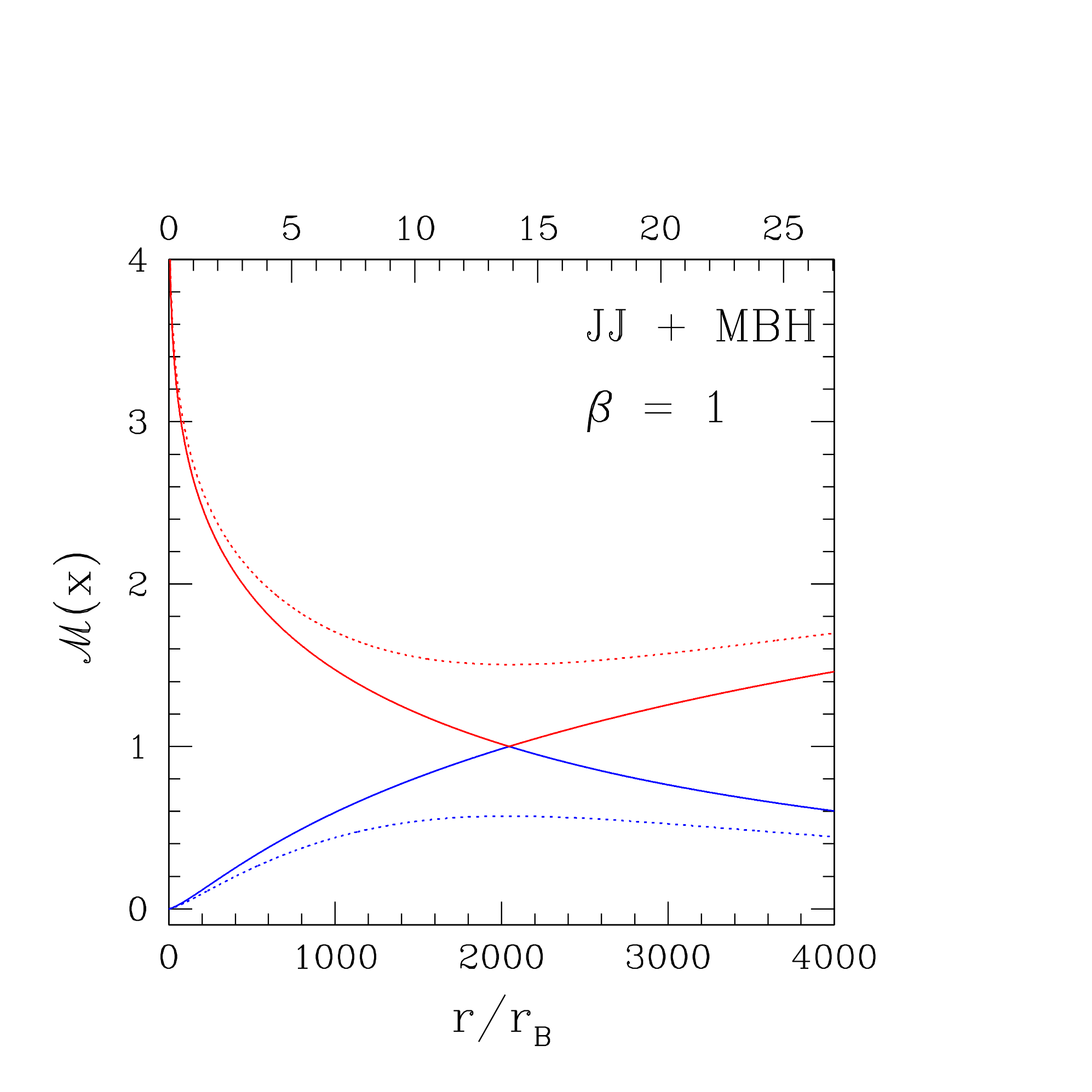}
\vskip -1.5truecm
\includegraphics[height=0.55\textwidth, width=0.55\textwidth]{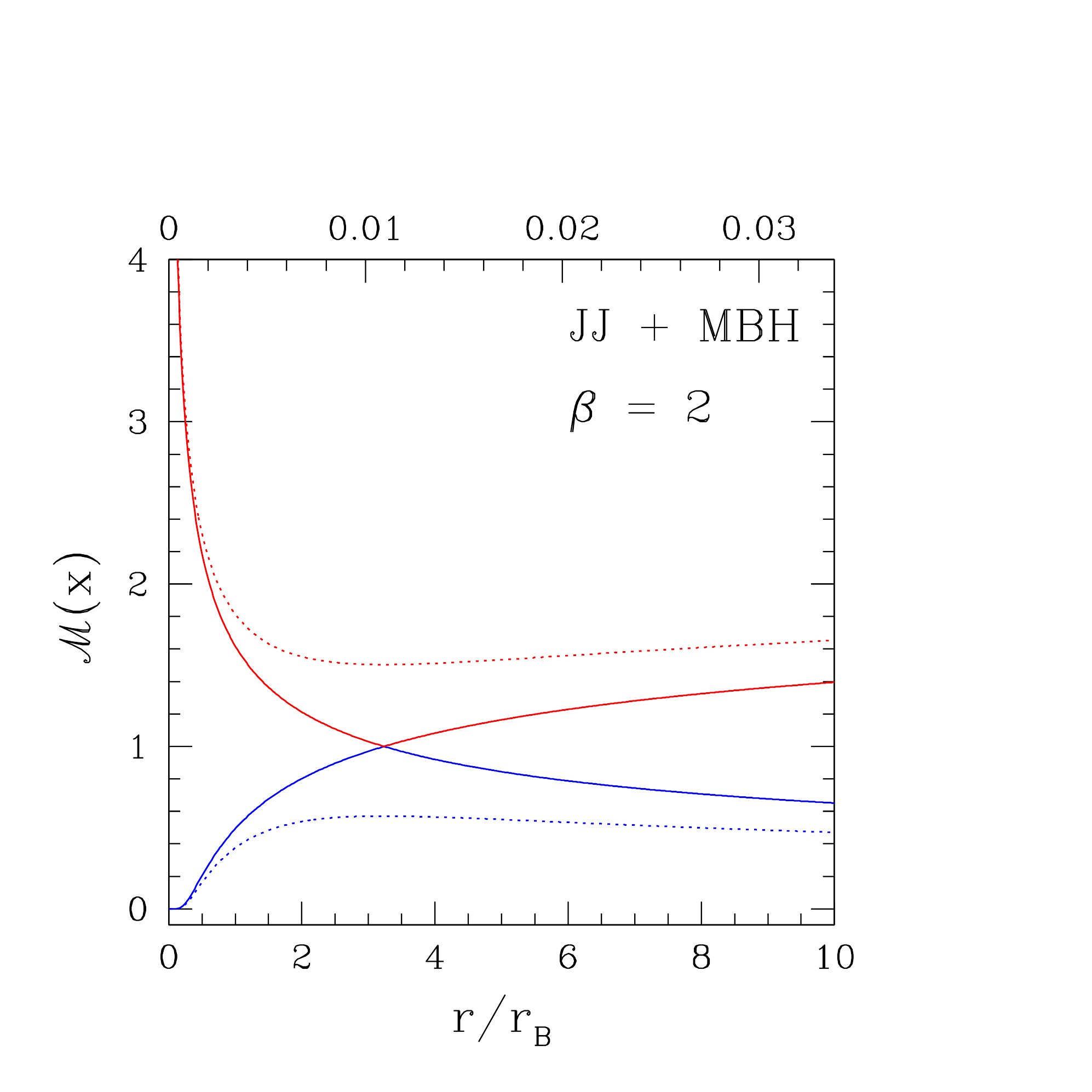}
\hskip -1truecm
\includegraphics[height=0.55\textwidth, width=0.55\textwidth]{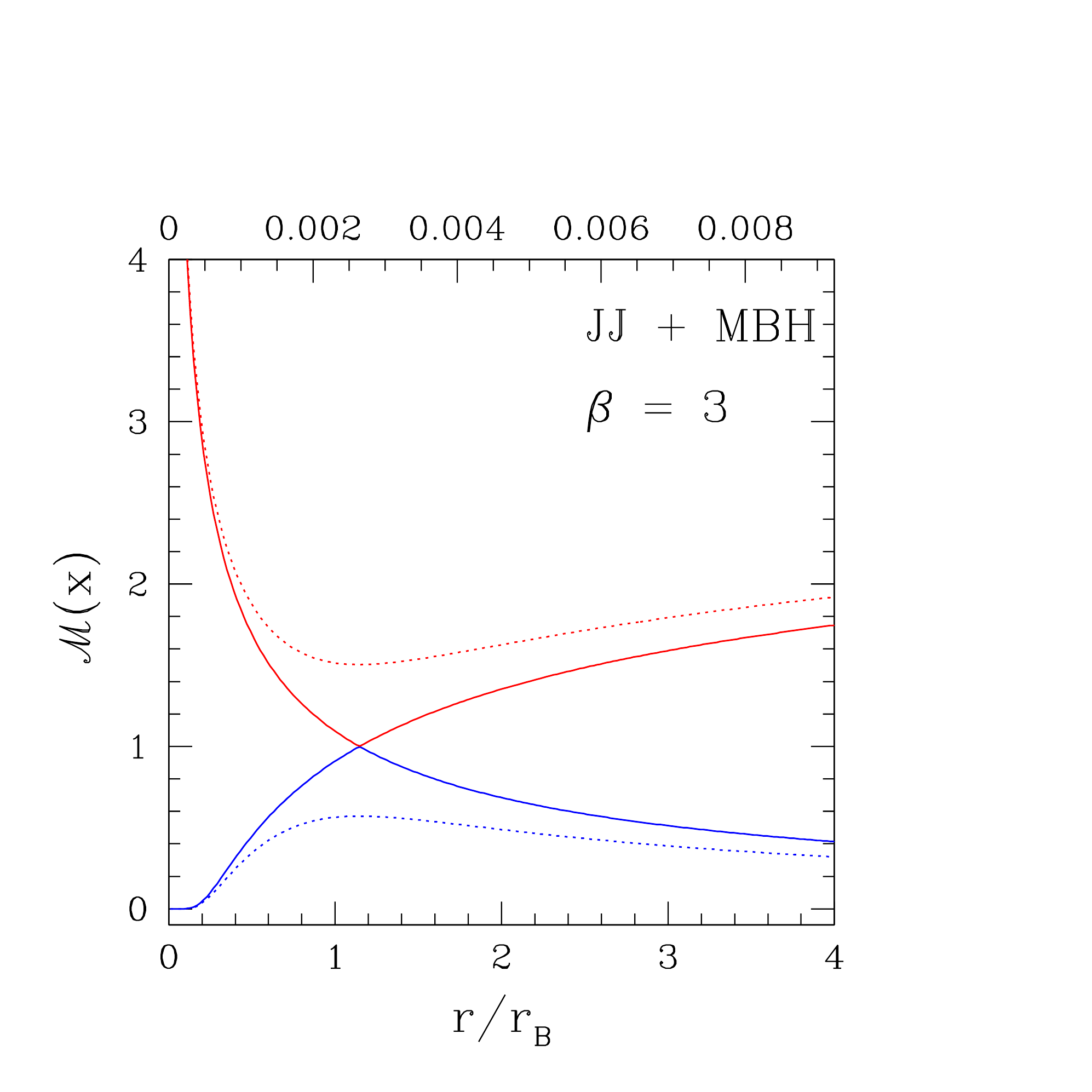}
%\vskip -1truecm
\caption{Mach number profiles as a function of $x=r/\rb$, for
  isothermal Bondi accretion (with $\chi=1$). The top left panel
  refers to classical Bondi accretion, and the other panels refer to
  accretion in a JJ galaxy model, plus a central MBH, with
  $\mu=2\times 10^{-3}$, $\MRg=\csig=20$, and $\beta =1, 2, 3$; the
  scale on the top gives the variable $r/\rs$.  The subsonic regime is
  plotted in blue, the supersonic one in red. Solid lines show the two
  critical solutions, dotted lines show two subcritical solutions
  (with $\lambda=0.8\lambdacr$ in the top left panel, and
  $\lambda=0.8\lambdat$ in the others).}
\label{}
\end{figure*}
%%%%%%%%%%%%%%%%%%%%%%%%%%%%%%

\subsection{Accretion in realistic galaxy models}

Here we show how to build JJ galaxy models with the main observed
properties of real galaxies, and how to derive the corresponding
parameters for isothermal accretion.

The first step consists of the determination of the stellar component
of the JJ galaxy.  This is done via the choice of two main galaxy
properties, for example the effective radius $\reff$ and the stellar
projected velocity dispersion $\siggp (0)$. For JJ models
$\siggp (0)=\sigg(0)$, and $\sigg (r)$ is quite flat at the center;
thus $\siggp (0)$ is very close to the emission-weighted projected
stellar velocity dispersion, within a small fraction of $\reff$ (as
typically given by observations).  For chosen $\reff$ and
$\siggp (0)$, then, one has $\rs\simeq 1.34 \reff$ [see below
eq. (20)], and then $\Ms$, from eqs. (32) and (21) once a value for
$\alpha\geq 1$ is fixed. The central MBH enters the problem via a
choice for $\mu$, that here we take to be $\mu=2\times 10^{-3}$ (e.g.,
Kormendy \& Ho 2013). Then the radius of influence $\Rinf$ can be
evaluated from eq. (37).  As an example, for a choice of $\reff=5$
kpc, and $\siggp (0)=210$ km s$^{-1}$, one has $\rs\simeq 6.7$ kpc,
$\alpha\Ms\simeq 1.38\times 10^{11}$ M$_{\odot}$, and
$\Rinf\simeq 4.6/\alpha$ pc, for a fiducial $\epsilon=0.5$.

The second step consists of the determination of the parameters $\MRg$
and $\csig$ that fix the total density distribution of the galaxy, and
in particular its total potential.  Since $\MRg = \alpha \csig$, we
can only fix either $\MRg$ or $\csig$.  It may be convenient to fix
$\csig$, for the following reason. A detailed dynamical modeling of
stellar kinematical data for galaxies of the local universe has shown
that the dark matter fraction within $\reff$ is low (e.g., Cappellari
2016). To fit with this result, one can use eq. (28), that relates
$\csig$ and the dark matter fraction within any radius $r$; thus, for
the desired (low) value of the dark matter fraction, $\csig$ remains
determined.  Figure 1 shows that the ratio $\MD(\reff)/\Mg(\reff)$ is
always in the range determined by the dynamical modeling, for
$\alpha=1$. The figure also shows that, for $\csig\ga 5$, the dark
matter fraction within $r=\reff$ is quite independent of $\csig$. This
means that a certain freedom remains in the choice of $\csig$, and we
can further constrain it from other considerations. If $\alpha=1$, we
may require that the scale-length of the dark halo ($r_{\rm NFW}$) is
larger than that of the stars (i.e., $\csih>1$).  For example a value
of the concentration parameter $c\simeq 10$, as predicted for galaxies
of the local universe (e.g., Dutton \& Macci\`o 2014), gives
$\csih=2.6$ for $\csig=20$ [eq. (30)].  Finally, one recovers the
stellar virial velocity dispersion
$\sigv^2=2 \calFg(\csig) \siggp^2(0)$, and then $\Tv$ from
$\sigv^2$. Since $\calFg(\csig)$ varies only by a factor of two for
$\csig=1$ to $\infty$, then in turn $\Tv$ varies at most by a factor
of two. For $\csig=20$, one has $\sigv\simeq 280$ km s$^{-1}$ and
$\Tv\simeq 2.0\times 10^6$ K (for $<\mu> =0.6$).

Having completely determined the galaxy structure with the choice of
two observed quantites, $\reff$ and $\siggp(0)$, and three parameters
($\mu$, $\alpha$, $\csig$), the next step consists in the
determination of the accretion properties. These are all fixed, once
the galaxy structure is fixed; only the gas temperature needs to be
chosen.  The first parameter describing accretion is $\MR$, obtained
from eq. (42).  The second parameter is $\xi$, that comes from
eq. (46), once the parameter $\beta$ is fixed in eq. (43). With the
choice of this last parameter, i.e., with the choice of $\Tinf$, all
the accretion properties are finally determined analytically, in
particular the gas sound speed $\csinf$ [eq. (43)], the Bondi radius
$\rb$ [eqs. (3) and (44)], the sonic radius $\rmin$ [eqs. (15) and
(47)], the critical accretion parameter $\lambdat$ [as described below
eq. (15)], and finally the Mach number profile $\calM$ [eq. (18)]. As
an example, for the galaxy model considered, for $\alpha=1$,
$\beta =1$ and $\csig =20$, one has $\MR=10^4$, $\rb\simeq 45$ pc,
$\xi\simeq 2.96\times 10^3$, $\rmin\simeq 93$ kpc, and
$\lambdat\simeq 5.23\times 10^7$. Instead, changing only the gas
temperature to $\beta =2$, one has $\MR$ unchanged, $\rb\simeq 23$ pc,
$\xi\simeq 5.91\times 10^3$, $\rmin\simeq 73$ pc, and
$\lambdat\simeq 2.85\times 10^6$.

Figure 5 shows the Mach number profiles for accretion on a MBH (the
classic Bondi problem), and on a MBH at the center of a JJ model, for
the minimum halo case and three values of $\beta$. For the three
galaxy models the top axis gives the radial scale in terms of $r/\rs$.
Again it is apparent how a modest increase in the gas temperature
produces a dramatic decrease of $\rmin$.  Figure 6 shows a comparison
between the gas velocity profile and the stellar velocity dispersion
profile, for the JJ models in Fig. 5.  Notice that near the center
$\sigBH\sim r^{-1/2}$ and $\calM \sim r^{-1/2}$ so that their ratio is
a constant; it can be easily shown that this constant is
${\sqrt{6\chi}}$, independently of $\alpha$, $\beta$, $\csig$, $\MRg$
and $\mu$. In principle, then, the value of $\sigBH$ close to the
center of a galaxy is a proxy for the (isothermal) gas inflow
velocity.

%%%%%%%%%%%%%%%%%%%%%%%%%%%%%%
\begin{figure}
\hskip -0.2truecm 
\includegraphics[height=0.55\textwidth, width=0.55\textwidth]{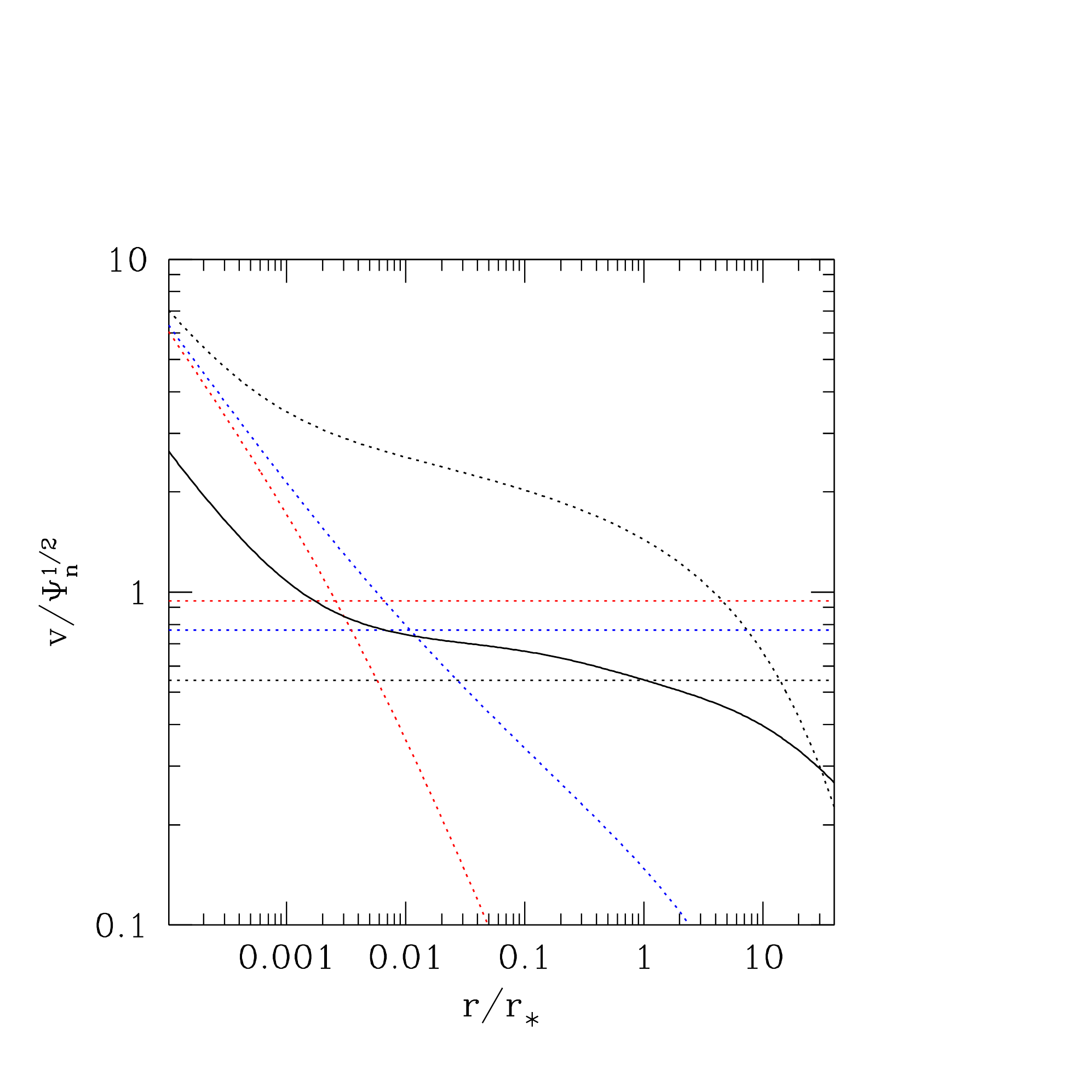}
\caption{Accretion velocity profile for the gas (dotted) and isotropic
  stellar velocity dispersion profile (solid), both normalized to
  $\sqrt{\Psin}$, for the minimum halo model with $\MRg=\csig=20$. The
  accretion solutions correspond to $\beta = 1,2,3$ and are given by
  the black, blue and red dotted lines, respectively. The horizontal
  dotted lines mark the corresponding values of $\csinf/\sqrt{\Psin}$.
  For each $\beta$ the intersection between the accretion velocity and
  the sound velocity marks the sonic point (bottom right panel in
  Fig. 2, solid dots).}
\label{}
\end{figure}
%%%%%%%%%%%%%%%%%%%%%%%%%%%%%%

\subsection{The bias in estimates of the mass accretion rate}  
\label{sec:biasclass}

We investigate here the use of the classical Bondi solution in
problems involving accretion on MBHs residing at the center of
galaxies. This use is common in the interpretation of observational
results, in numerical investigations, or in cosmological simulations
(see Sect. 1). In many such studies, when the instrumental resolution
is limited, or the numerical resolution is inadequate, an estimate of
the mass accretion rate is derived using the classical Bondi solution,
taking values of temperature and density measured at some finite
distance from the MBH. This procedure clearly produces an estimate
that can depart from the true value, even when assuming that accretion
fulfills the hypotheses of the Bondi model (stationariety, spherical
symmetry, etc.).  KCP16 developed the analytical set up of the problem
for generic polytropic accretion, with the inclusion of the effects of
radiation pressure and of a galactic potential; they also investigated
numerically the size of the deviation for Hernquist galaxies. CP17
presented a detailed exploration of the deviation for isothermal
accretion in one-component Jaffe and Hernquist galaxies was done by
CP17.  Here we consider the more realistic case of two-component JJ
models, in the isothermal case, exploiting the fully analytical
character of JJ models.

We first consider the deviation of the estimate of the mass accretion
rate for the classical Bondi solution, when taking values of
temperature and density measured at some finite distance from the MBH.
For assigned values of $\rhoinf$, $\Tinf$, $\gamma$ and $\Mbh$, the
Bondi radius $\rb$ and the critical accretion rate $\Mdotb$ are given
by eq. (3) and by eq. (5) with $\lambda=\lambdacr$.  If one inserts in
these equations the values of $\rho (r) $ and $T(r)$ at a finite
distance $r$ from the MBH, taken along the classical Bondi solution,
and considers them as ``proxies'' for $\rhoinf$ and $\Tinf$, then an
{\it estimated} value for the accretion radius ($\rbe$) and mass
accretion rate ($\Mdotbe$) is obtained:
\begin{equation} 
\rbe (r)\equiv {G\Mbh\over\cs^2(r)},\quad
\Mdotbe(r) \equiv 4 \pi \rbe^2(r)  \lambdacr \rho(r)\cs (r).
\end{equation}
The question is how much $\rbe$ and $\Mdotbe$ depart from the true
values $\rb$ and $\Mdotb$, as a function of $r$.  In the isothermal
case the sound speed is constant, with $\cs (r) =\csinf$, and then
$\rbe (r)=\rb$, independently of the distance from the center at which
the temperature is evaluated. Then
$\Mdotbe(r) = 4 \pi \rb^2 \lambdacr \rho(r)\csinf$; at infinity,
$\Mdotbe = \Mdotb$.  At finite radii, instead
\begin{equation}
{\Mdotbe(r)\over \Mdotb} =\rhotil(x)={\lambdacr\over x^2 \calM(x)},
\end{equation}
where the last identity comes from eq. (9), and $\calM(x)$ is given in
eq. (19) of CP17.  The deviation of $\Mdotbe$ from $\Mdotb$ then is
just given by ${\rhotil(x)} $ at the radius $r$ where the ``measure''
is taken.  Thus, $\Mdotbe$ gives an overestimate of $\Mdotb$, and this
overestimate becomes larger for decreasing $x$ (see Fig. 1 in KCP16,
and Fig. 4 in CP17).

%%%%%%%%%%%%%%%%%%%%%%%%%%%%%%
\begin{figure*}
\hskip -0.2truecm  
\includegraphics[height=0.55\textwidth, width=0.55\textwidth]{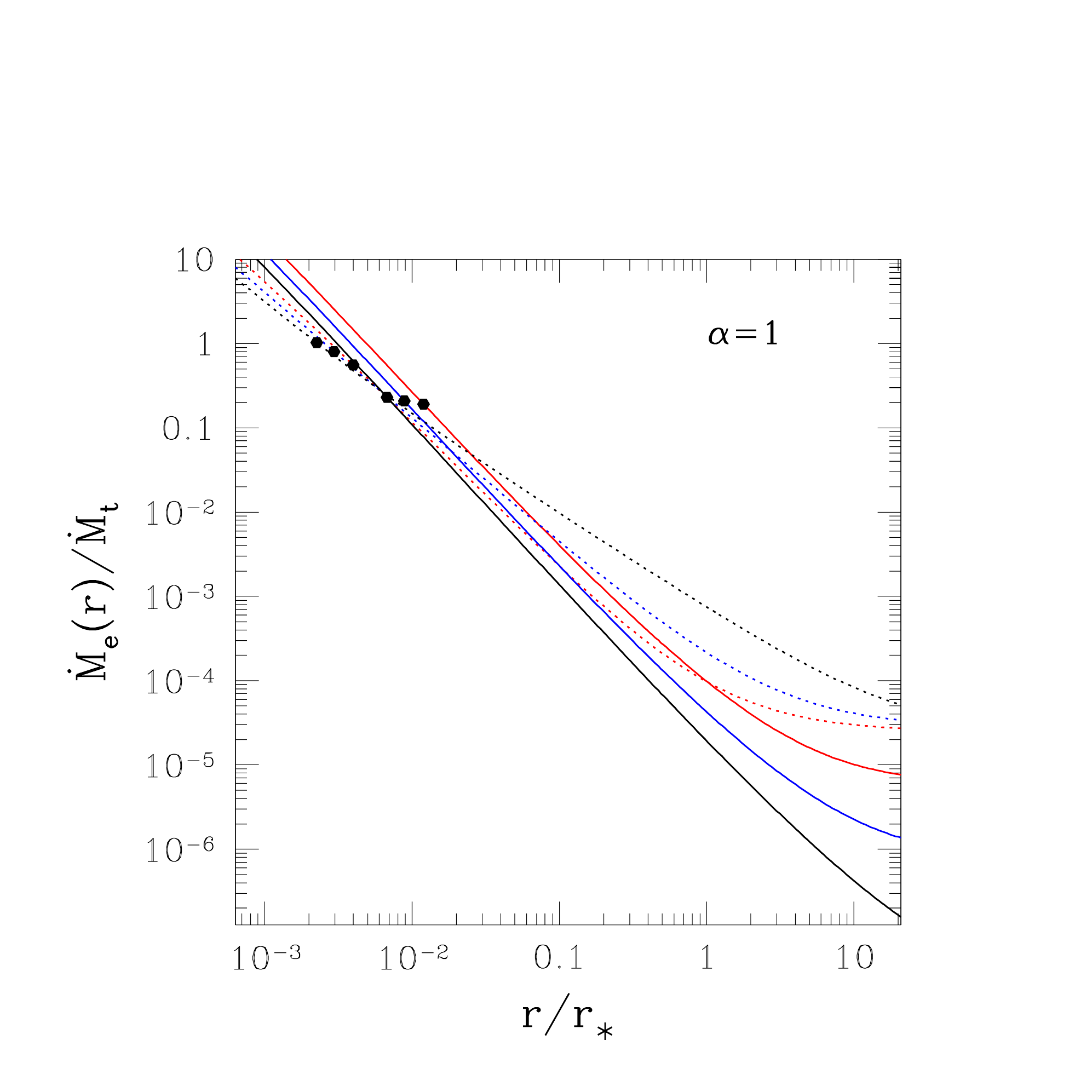}
\includegraphics[height=0.55\textwidth, width=0.55\textwidth]{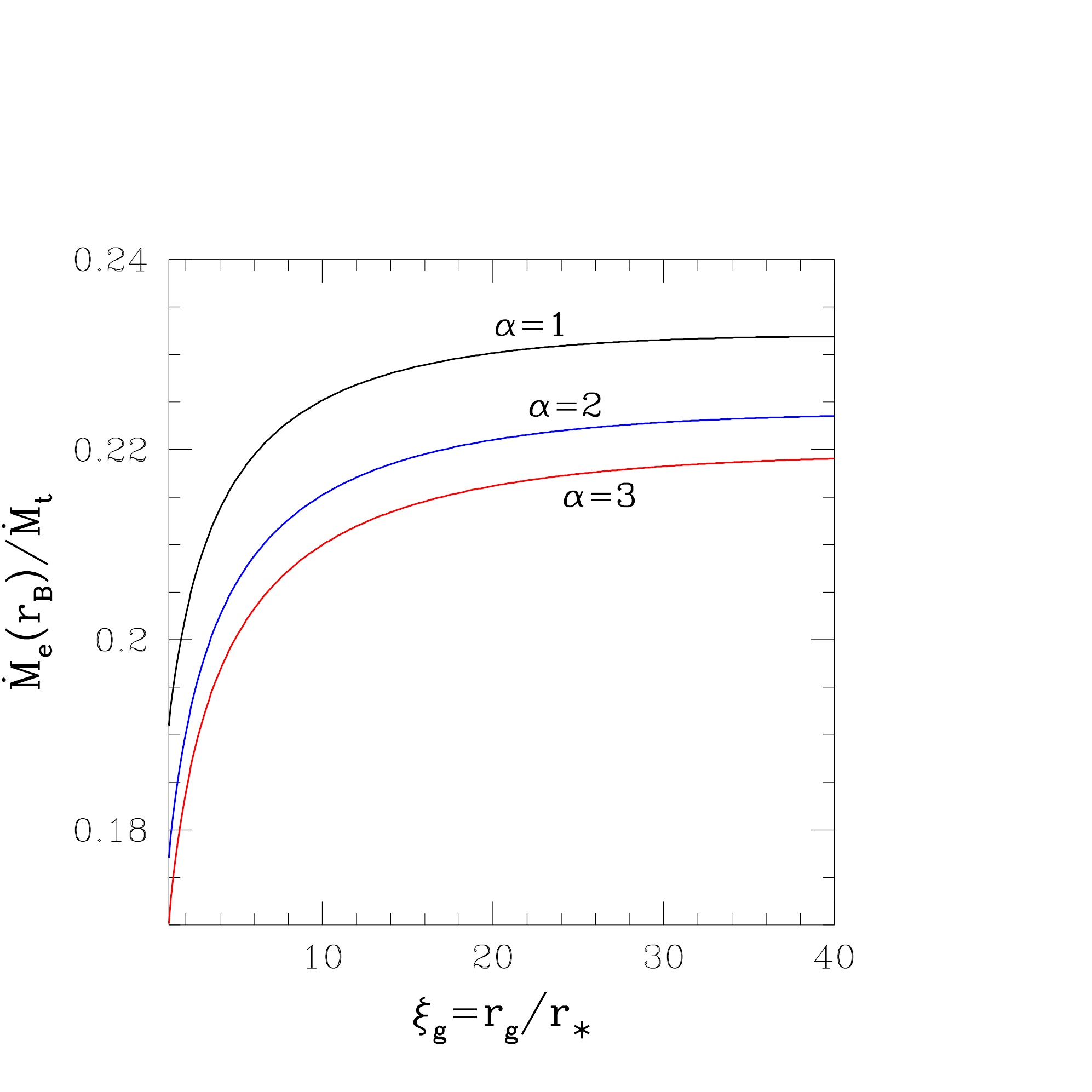}\\
\caption{Left: ratio between the estimate of the accretion rate
  $\Mdotbe$ and the true accretion rate $\Mdott$, from eq. (52), as a
  function of $r/\rs$, in the minimun halo case ($\alpha =1$), for
  $\chi =1$, $\beta =1$, and $\csig=1$ (red), $\csig=3$ (blue), and
  $\csig=20$ (black). The dotted lines correspond to $\beta =3$, and
  the solid dots mark the position of the Bondi radius $\rb$ in
  eq. (44).  Right: the ratio $\Mdotbe/\Mdott$ evaluated at $\rb$, as
  a function of $\csig$, for three values of $\alpha$, and $\chi =1$,
  $\beta =1$. In both panels $\mu=0.002$. }
\label{}
\end{figure*}
%%%%%%%%%%%%%%%%%%%%%%%%%%%%%%

In presence of a galaxy, the departure of $\Mdotbe(r)$ of eq. (50)
from the true mass accretion rate $\Mdott =4 \pi \rb^2 \lambdat
\rhoinf\csinf $, is:
\begin{equation}
{\Mdotbe(r)\over\Mdott}={\lambdacr\rhotil(x)\over\lambdat}={\lambdacr\over
  x^2\calM(x)},
\end{equation}
where $\rho (r) $ is taken along the solution for accretion within the
potential of the galaxy\footnote{ In the monoatomic adiabatic case
  $\gamma=5/3$, one has that $\Mdotbe (r)=\lambdacr\Mdott/\lambdat$
  independently of the distance from the center, while $\rbe(r)$
  departs from $\rb$ (KCP16, eq. 42).}, the last identity comes from
eq. (19), and $\calM(x)$ is given in eq. (18).

Figure 7 (left panel) shows the trend of $\Mdotbe/\Mdott$ with $r$.
One sees that the use of $\rho(r)$ instead of $\rhoinf$ leads to an
overestimate for $r$ taken in the central regions, while $\Mdotbe(r)$
becomes an {\it underestimate} for $r\ga$ a few$\times
10^{-3}\rs$. The radius marking the transition from the region in
which there in an overestimate, to that where there is an
underestimate, depends on the specific galaxy model.  An important
consequence of the results in Fig. 7 is that, for example, in
numerical simulations not resolving $\rb$, $\Mdotbe$ should be boosted
by a large factor to approximate the true accretion rate
$\Mdott$. Moreover, since $\Mdotbe/\Mdott$ increases steeply with $r$,
this ``boost factor'' in turn also varies steeply with $r$. In the
right panel of Fig. 7 we also show the bias measured at the Bondi
radius as a function of $\csig$. The panel indicates an underestimate
by roughly a factor of 5.

It is instructive to find the reason for the trend of $\Mdotbe$ near
the center and at large radii.  From eq. (52) and the expansion of
$\calM$ for $x\to 0^+$ and for $x \to \infty$, one has:
\begin{equation}
{\Mdotbe(r)\over\Mdott}\sim {\lambdacr\over\sqrt{2\chi}x^{3/2}}, \quad  x \to 0^+,
\end{equation}
and
\begin{equation} {\Mdotbe(r)\over\Mdott}\sim {\lambdacr\over\lambdat},
 \quad x \to \infty.
\end{equation}
Therefore near the center $\Mdotbe/\Mdott\sim r^{-3/2}$, while at
large radii, as in general $\lambdat>>\lambdacr$ (see Fig. 4),
$\Mdotbe/\Mdott$ becomes very small.

\section{Summary and conclusions}

The classical Bondi accretion theory is the tool commonly adopted in
many investigations where an estimate of the accretion radius and the
mass accretion rate is needed. In this paper, extending the results of
previous works (KCP16, CP17), we focus on the case of isothermal
accretion in two-component galaxies with a central MBH, and with
radiation pressure contributed by electron scattering in the optically
thin regime.  In CP17 it was shown that the radial profile of the Mach
number, and the critical eigenvalue of the isothermal accretion
problem, can be expressed analitycally in Jaffe and Hernquist
potentials with a central MBH. Here we adopt the two-component JJ
galaxy models presented in CZ18. These are made of a Jaffe stellar
component plus a dark halo such that the total density is also
described by a Jaffe profile; all the relevant dynamical properties of
JJ models, including the solution of the Jeans equations for the
stellar component, can be expressed analytically.  Therefore, the
results of CP17 and CZ18 give the opportunity of building a family of
two-component galaxy models where all the accretion properties can be
given analytically, and then explored in detail, with no need to
resort to numerical studies.  The main results of this work can be
summarized as follows.

1) The parameters describing accretion in the hydrodynamical solution
of CP17 ($\MR$ and $\xi$) have been linked to the galaxy
structure. In particular, it is assumed that the isothermal gas has a
temperature $\Tinf$ proportional to the virial temperature of the
stellar component, $\Tv$. Then, simple formulas are derived relating
the galactic properties (as the effective radius, $\reff$, and the
radius of influence of the MBH, $\Rinf$) with those describing
accretion (as the Bondi radius $\rb$, and the sonic radius $\rmin$).
The critical accretion parameter $\lambdat$ is also expressed as a
function of the galactic properties.

2) For realistic galaxy structures, $\rb$ is of the order of a
few$\times 10^{-3}\rs$, and $\Rinf$ is of the order of
$\simeq 0.1\rb$. For $\Tinf =\Tv$, the sonic radius $\rmin$ is of the
order of a few $\reff$. For moderately higher values of $\Tinf$,
$\rmin$ suddenly drops to radii within $\rb$. The same happens also
for a small increase of the polytropic index above unity, and this
behavior is reminiscent of the similar jump shown by $\rmin$ in
Hernquist models, as discussed in CP17.  As a consequence, accretion
in JJ models can switch from being supersonic over almost the whole
galaxy to being everywhere subsonic, except for $r\la\rb$. An
explanation for this phenomenon is given.

3) As for the isothermal accretion in one-component Jaffe models, the
determination of the critical accretion parameter involves the
solution of a quadratic equation, and there is only one sonic point
for any choice of the parameters describing the galaxy.  In presence
of the galaxy, $\lambdat$ is several orders of magnitude larger than
without the galaxy. It is found that Bondi accretion in JJ models in
absence of a central MBH (or when $\chi =0$) is possible, provided
that $\Tinf$ is lower than a critical value and we derive the esplicit
formula for it. This critical value depends only on $\csig$, and is in
the range $3/2\leq\Tinf/\Tv\leq 3$. It also determines the the jump in
$\rmin$ in models with the central MBH.

4) We provide a few examples of accretion in realistic galaxy models,
and present the resulting Mach number profiles, the trends of the
accretion velocity and of the isotropic stellar velocity dispersion
profiles.

5) We finally examine the problem of the deviation from the true value
$\Mdott$ of an estimate of the mass accretion rate $\Mdotbe(r)$
obtained adopting the classical Bondi solution for accretion on a MBH,
where the gas density and temperature at some finite distance from the
center are inserted, as proxies for their values at infinity.  The
size of the departure of $\Mdotbe(r)$ from $\Mdott$, that is
determined by the presence of the galaxy, is given as a function of
the distance $r$ from the center. $\Mdotbe(r)$ {\it overestimates}
$\Mdott$, if the gas density is taken in the very central regions, and
{\it underestimates} $\Mdott$ if it is taken outside a few Bondi
radii. This shows how sensitive to the model parameters is the
determination of a physically based value for the so-called ``boost
factor'' adopted in simulations, and that in general a universally
valid prescription is impossible.

\begin{acknowledgements}

\end{acknowledgements}

\end{document}